\documentclass[twocolumn,prd,nofootinbib,superscriptaddress]{revtex4}
\usepackage{hyperref}

\newcommand\ForInternalReference[1]{}
\newcommand\SkipForEarlyCirculation[1]{}

\newcommand\SkipPP[1]{}
\usepackage{verbatim}
\usepackage{graphicx}
\usepackage{dcolumn}
\usepackage{bm}
\usepackage{color}
\usepackage{xspace}
\usepackage{url}
\usepackage{amsmath}
\usepackage{float}
\usepackage{multirow}
\usepackage{times}
\newcommand\optional[1]{}

\newcommand\qmstateproduct[2]{\left\langle#1|#2\right\rangle}

\newcommand\unit[1]{{\rm #1}}

\newcommand\editremark[1]{{\color{red} #1}}
\usepackage{color}
\definecolor{amber}{rgb}{1.0, 0.75, 0.0}
\definecolor{orange}{rgb}{1.0, 0.5, 0.0}
\definecolor{amaranth}{rgb}{0.9, 0.17, 0.31}

\graphicspath{{./figures/}}
\newcommand{\mc}{{\cal M}}

\def\ltsima{$\; \buildrel < \over \sim \;$}
\def\simlt{\lower.5ex\hbox{\ltsima}}
\def\gtsima{$\; \buildrel > \over \sim \;$}
\def\simgt{\lower.5ex\hbox{\gtsima}}

\newcommand{\IMRPD}{\textsc{IMRPhenomD}\xspace}

\newcommand{\SEOBA}{\textsc{SEOBNRv4}\xspace}

\newcommand\RIFT{RIFT}

\newcommand{\RIT}{\affiliation{Center for Computational Relativity and Gravitation, Rochester Institute of Technology, Rochester, New York 14623, USA}}
\newcommand{\ICERM}{\affiliation{Institute of Computational and Experimental Research in Mathematics, Brown University, Rhode Island 02903, USA}}
\begin{document}

%\title{Waveform systematics of compact binary coalescence models}
\title{Assessing and marginalizing over compact binary coalescence waveform systematics with RIFT}
\author{A.~Z.~Jan}
\author{A.~B.~Yelikar}
\RIT
\author{J.~Lange}
\ICERM
\RIT
\author{R.~O'Shaughnessy}
\RIT

%% PLAN
%% * (a) model averaging at lower cost
%% * (b) predict biases
%% * (c) propagation into population error, which we show is substantial  (for SPIN)

\begin{abstract}
As Einstein's equations for binary compact  object inspiral have only been approximately or intermittently solved by
analytic or numerical methods, the models used to infer parameters of gravitational wave (GW) sources are subject to waveform
modeling uncertainty.  Using a simple scenario, we illustrate these differences, then introduce a very efficient
technique to marginalize over waveform uncertainties, relative to a pre-specified sequence of waveform models.
Being based on RIFT, a very efficient parameter inference engine, our technique can directly account for any available
models, including very accurate
but computationally costly waveforms.  Our evidence- and likelihood-based method works robustly on a point-by-point
basis, enabling accurate marginalization for models with strongly disjoint posteriors while simultaneously   increasing the
re-usability and efficiency of our intermediate calculations.
%% This project aims to assess the nature and impact of systematic differences between different 
%% waveform approximations for binary black hole, binary neutron star, and black hole/neutron star mergers.
%% \textcolor{red}{More content needed!!}
\end{abstract}
\maketitle

\section{Introduction}
\label{sec:Intro}

% Boilerplate
%  - GW observatoins
%  - systematic error: exemplified with differences seen for 190521g, 190412m, and forthcoming GWTC-1
%  - waveform accuracy studies show it will still matter in the near future (previous work) ...
Since the first gravitational wave detection GW150914 \cite{DiscoveryPaper}, the Advanced Laser Interferometer Gravitational Wave Observatory (LIGO) 
~\cite{2015CQGra..32g4001L} and  Virgo
\cite{gw-detectors-Virgo-original-preferred,TheVirgo:2014hva}
detectors have
continued to discover gravitational waves (GW) from coalescing binary black holes (BBHs) and neutron stars.
The properties of each source are inferred by comparing each observation to some estimate(s) for the GW emitted when a BBH
merge, commonly called an approximant.  
As illustrated most recently by GW190521 \cite{LIGO-O3-GW190521-discovery,LIGO-O3-GW190521-implications},
GW190814 \cite{LIGO-O3-GW190814}, GW190412 \cite{LIGO-O3-GW190412}, and the discussion in GWTC-2 \cite{LIGO-O3-O3a-catalog}, these
approximations disagree more than enough to produce noticable differences, consistent with prior work 
\cite{Shaik:2019dym,gwastro-Systematics-Williamson2017,Purrer:2019jcp}.
Despite ongoing generation of new waveforms with increased accuracy \cite{gwastro-mergers-IMRPhenomP,gwastro-mergers-IMRPhenomPv3,gwastro-SEOBNRv4,2019arXiv190509300V,gwastro-mergers-IMRPhenomXP,2020PhRvD.102d4055O}, these previous investigations suggest that waveform model systematics
can remain a limiting factor in inferences about individual events \cite{Shaik:2019dym} and populations 
\cite{gwastro-PopulationReconstruct-Parametric-Wysocki2018,Purrer:2019jcp}.

% Previous related work
%  - Ashton and Khan: optimal method to marginalize over waveform uncertainty
%  - Challenge of waveform cost: can't perform calculation for costly waveforms at all with conventinoal methods (e.g., 190412m)
% Goal of this paper:
%   - deploy equivalent, but much more efficient and extendable, performed with RIFT.
%     Can use very high-precision MC integration, and RIFT highly scalable so usable with costly approximations
%   - use simple scenario to assess waveform accuracy, and show how it can be mitigated by our method

% Paper organization
Recently, Ashton and Khan \cite{Ashton:2019leq} described and illustrated marginalizing between a discrete set of
waveform models in a fully Bayesian way.  In this procedure, the waveform-marginalized posterior is the  weighted
average of the posteriors $p_k(\theta)$ derived from each
 waveform model $k$ alone, weighted by the evidence $Z_k$  for (and prior $p_k$ for) each model $k$: $p(\theta) = [\sum_k
   p_k(\theta)p_kZ_k]/\sum_qp_q Z_q$.   This extremely simple procedure faces one obvious limitation: analysis must be
 performed for every waveform model  of interest.   Unfortunately, as many of the most accurate time-domain waveform models
 incur exceptionally high evaluation costs, and as most conventional parameter estimation (PE) engines like LALInference \cite{Veitch:2014wba}
 or BILBY \cite{2019ApJS..241...27A} are limited
 by this cost, the universe of possible waveforms must often omit the most expensive and accurate waveform models.
As the RIFT parameter inference engine  circumvents several issues associated with waveform evaluation cost
\cite{gwastro-PENR-RIFT,gwastro-PENR-RIFT-GPU}, despite retaining the original waveform implementation (i.e., no
surrogate generation), in this work
we examine novel extensions of this waveform-marginalization technique which are uniquely adapted to RIFT's algorithm. 
Using a simple toy model, we demonstrate the pernicious effects of model systematics, then show how our technique
efficiently mitigates them.

%OUTLINE

%Method section:

%review of RIFT
%Review of waveform models, and prior work on comparing them
%PP plots and fiducial populations used in this work: what range of parameters are you exploring
%zero noise runs used to identify the secular effects
%introduce diagnostics - how to you tell if posteriors are different, and what is "different enough" : shifts, etc. PP plots . BF
%Results 1 section: Demonstrating the impact of model differences

%Fiducial PP plots for the two models. (Reference prior work). Fig 1! Identify that this population shows differences
%other ways to characterize those differences
%can we tell which model is correct from the data? Discussing BF ... not high enough to distinguish which is the correct model
%(pending section) this produces a biased population recovery
%offsets. Attempts to fit offsets
%Methods/results 2: Mitigating the offsets:

%describe the marginalization method, see appendix
%describe the PP test for our model marginalization
%Plot showing it works
%discuss merits of this method

%Optional : try precessing binaries (harder)

% Need to add more background information

This paper is organized as follows.
In Section \ref{sec:methods}, we review the use of RIFT for parameter inference; the two waveform models used in this
work; the use of probability-probability (PP) plots to diagnose systematic error with noise; the use of zero-noise PE to isolate the systematic
uncertainty between waveforms; and our waveform marginalization technique.
In Section \ref{sec:results}, we use two well-studied waveform models to demonstrate the impact of contemporary model
systematics, then marginalize over them.  We emphasize that all calculations in this section adopt signal amplitudes and
masses consistent with current observations.  
In Section \ref{sec:conclude}, we summarize our results and discuss their potential applications to future GW source and
population inference.

\section{Methods}
\label{sec:methods}

\subsection{RIFT review}
\label{subsec:RIFT}

A coalescing compact binary in a quasicircular orbit can be completely characterized by its intrinsic
and extrinsic parameters.  By intrinsic parameters we refer to the binary's  masses $m_i$, spins, and any quantities
characterizing matter in the system.  For simplicity and reduced computational overhead, in this work we assume all
compact object spins are aligned with the orbital angular momentum. 
By extrinsic parameters we refer to the seven numbers needed to characterize its spacetime location and orientation.  
We will express masses in solar mass units and
 dimensionless nonprecessing spins in terms of cartesian components aligned with the orbital angular momentum
 $\chi_{i,z}$.   We will use $\lambda,\theta$ to
refer to intrinsic and extrinsic parameters, respectively.

RIFT \cite{gwastro-PENR-RIFT}
consists of a two-stage iterative process to interpret gravitational wave data $d$ via comparison to
predicted gravitational wave signals $h(\lambda, \theta)$.   In one stage, for each  $\lambda_\alpha$ from some proposed
``grid'' $\alpha=1,2,\ldots N$ of candidate parameters, RIFT computes a marginal likelihood 
\begin{equation}
 {\cal L}_{\rm marg}\equiv\int  {\cal L}(\bm{\lambda} ,\theta )p(\theta )d\theta
\end{equation}
from the likelihood ${\cal L}(\bm{\lambda} ,\theta ) $ of the gravitational wave signal in the multi-detector network,
accounting for detector response; see the RIFT paper for a more detailed specification.  
In the second stage,  RIFT performs two tasks.  First, it generates an approximation to ${\cal L}(\lambda)$ based on its
accumulated archived knowledge of marginal likleihood evaluations 
$(\lambda_\alpha,{\cal L}_\alpha)$.  This approximation can be generated by gaussian processes, random forests, or other
suitable approximation techniques.   Second, using this approximation, it generates the (detector-frame) posterior distribution
\begin{equation}
\label{eq:post}
p_{\rm post}=\frac{{\cal L}_{\rm marg}(\bm{\lambda} )p(\bm{\lambda})}{\int d\bm{\lambda} {\cal L}_{\rm marg}(\bm{\lambda} ) p(\bm{\lambda} )}.
\end{equation}
where prior $p(\bm{\lambda})$ is the prior on intrinsic parameters like mass and spin.    The posterior is produced by
performing a Monte Carlo integral:  the evaluation points and weights in that integral are weighted posterior samples,
which are fairly resampled to generate conventional independent, identically-distributed ``posterior samples.''
For further details on RIFT's technical underpinnings and performance,   see
\cite{gwastro-PENR-RIFT,gwastro-PENR-RIFT-GPU,gwastro-mergers-nr-LangePhD}.

\subsection{Waveform models}

In this work, we employ two well-studied models for non-precessing binaries, whose differences are  known to be
significant. We use \SEOBA\cite{gwastro-SEOBNRv4}, an effective-one-body model for quasi-circular inspiral, and  \IMRPD
\cite{IMRDI,IMRDII}, a phenomenological frequency-domain inspiral-merger-ringdown model.
%it's surrogate frequency-domain version SEOBNRv4\_ROM \cite{gwastro-SEOBNRv4},

%\editremark{below copied from RIFT paper -- too verbose}
The effective-one-body (EOB) approach models the inspiral and spin dynamics
of coalescing binaries via an ansatz for the two-body Hamiltonian~\cite{gw-astro-EOBspin-Tarrachini2012}, 
whose corresponding equations of motion are numerically solved in the time
domain.
For non-precessing binaries, outgoing gravitational
radiation during the inspiral phase is generated using an ansatz for resumming the post-Newtonian expressions for
outgoing radiation including non-quasicircular corrections, for the leading-order $\ell=2$ subspace.  For the  merger phase of 
non-precessing binaries,  the gravitational radiation is generated via a resummation of many quasinormal modes, with coefficients 
chosen to ensure smoothness. The final BH's mass and spin, as well as some parameters in the non-precessing inspiral model, 
are generated via calibration to numerical relativity simulations of BBH mergers.

The  \IMRPD model is a part of an approach that attempts to  approximate the leading-order ($\ell=2$) gravitational wave radiation 
using phenomenological fits
to the Fourier transform of the gravitational wave strain, computed from numerical relativity simulations and post-newtonian
calculation~\cite{nr-Jena-nonspinning-templates2007, gwastro-nr-Phenom-Lucia2010, gwastro-mergers-IMRPhenomP}.  
Also using information
about the final BH state, this phenomenological frequency-domain approach matches standard approximations for
the post-Newtonian gravitational wave phase to an approximate, theoretically-motivated spectrum characterizing merger
and ringdown.

\subsection{Fiducial synthetic sources and PP tests}
\label{sec:sub:pop}

\begin{figure}
\includegraphics[scale=0.4]{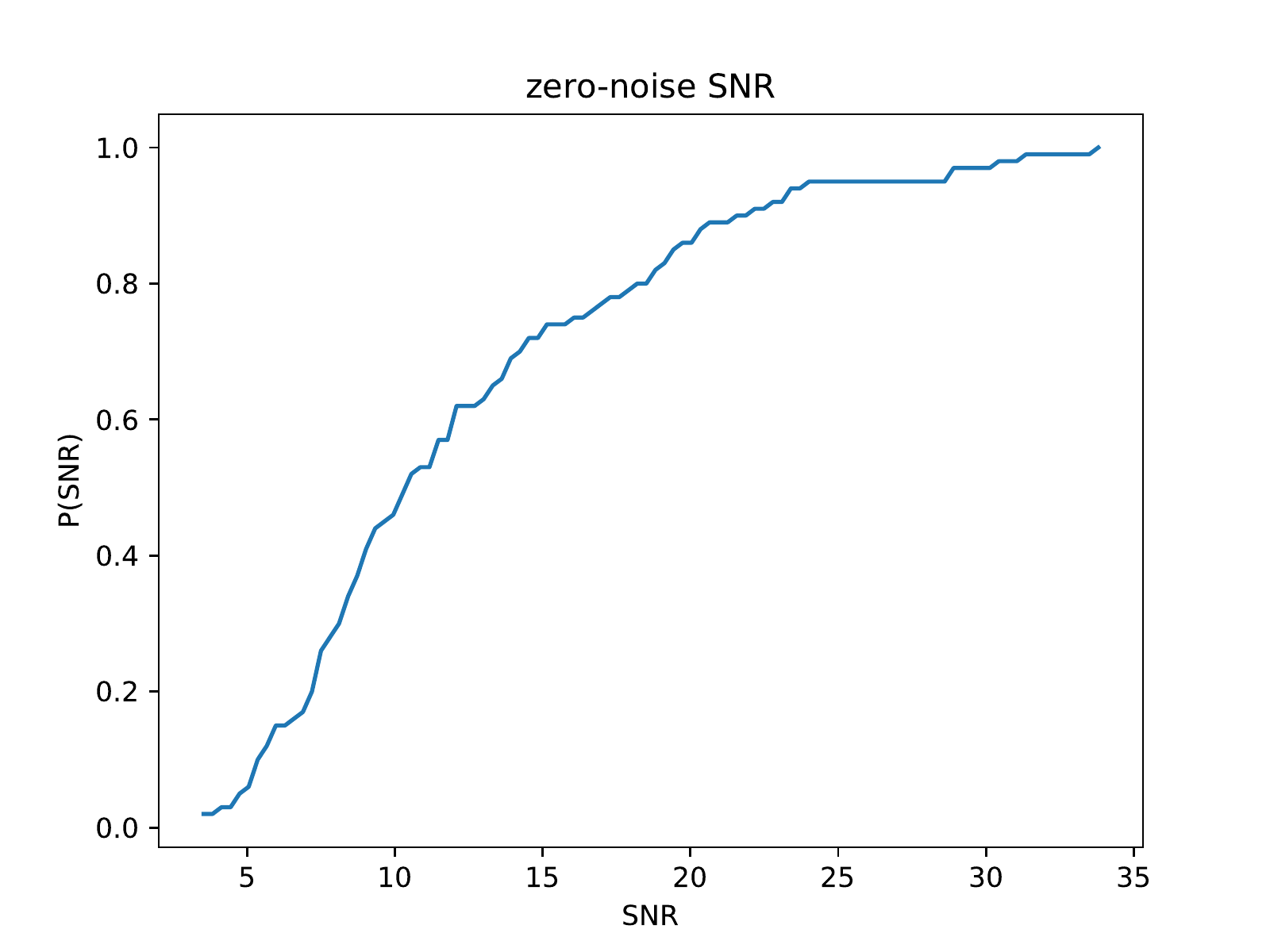}
\caption{Cumulative SNR distribution for a synthetic population of 100 events drawn from the fiducial BBH population described
  in Section \ref{sec:sub:pop}.  To avoid ambiguity, this figure shows the expected SNR (i.e., the SNR evaluated using a
zero-noise realization). }
\label{SNR-CDF}
\end{figure}

We will only explore the impact of systematics over a limited fiducial population.
Specifically, we consider a universe of synthetic signals for 
3-detector networks, with masses
drawn uniformly in $m_i$ in the region bounded by $\mc/M_\odot \in [30,60 ]$ and $\eta \in [0.2, 0.25]$ and with
extrinsic parameters drawn uniformly in sky position and isotropically in Euler angles, with source luminosity
distances  drawn proportional to $d_L^2$  between
$1.5\unit{Gpc}$ and $4\unit{Gpc}$. 
These bounds are expressed in terms of $\mc=(m_1m_2)^{3/5}/(m_1+m_2)^{1/5}$ and $\eta=m_1m_2/(m_1+m_2)^2$, 
and encompass the detector-frame parameters of many massive binary black holes seen in GWTC-1 \cite{LIGO-O2-Catalog} and
GWTC-2 \cite{LIGO-O3-O3a-catalog}.
All our sources have non-precessing spins, with each component assumed to be uniform
between $[-1,1]$.
%\editremark{correct above depending on wha was actually done}
(For complete reproducibility, we use \texttt{\SEOBA} and \texttt{ \IMRPD}, starting the signal evolution at $18\unit{Hz}$ but the
likelihood integration at $20\unit{Hz}$, performing all analysis with $4096\unit{Hz}$ timeseries in Gaussian noise with
known advanced LIGO design PSDs \cite{LIGO-aLIGODesign-Sensitivity-Updated}.  For each
synthetic event and for each interferometer, the same noise realization is used for both waveform approximations.
Ensuring convergence of the analyses, the differences between them therefore arise solely due to waveform systematics. 
For context, Figure  \ref{SNR-CDF} shows the  cumulative SNR distribution of one specific synthetic population
generated from this distribution. Though a small fraction have substantial signal amplitudes, most events are near or
below the level of typical detecton candidates.  By using a very modest-amplitude population to assess the impact of
waveform systematics, we demonstrate their immediate impact on the kinds of analyses currently being performed on real
observations, let alone future studies.
%% . This provides more weight to
%%  our study as we expect most GW events to be sub-threshold events, and even for those events, we see significant differences between waveform models. \\

One way to assess the performance of parameter inference is a probability-probability plots (usually denoted PP plot) \cite{mm-stats-PP}.
  Using \RIFT{} on each source $k$, with true parameters $\mathbf{\lambda}_k$, we estimate
the fraction of the posterior distributions which is below the true source value $\lambda_{k,\alpha}$   [$\hat{P}_{k,\alpha}(<\lambda_{k,\alpha})$] 
for each intrinsic parameter $\alpha$, again assuming all sources
have zero spin.  After reindexing the sources so $\hat{P}_{k,\alpha}(\lambda_{k,\alpha})$ increases with $k$ for some
fixed $\alpha$, the top panel of Figure \ref{pp-IMRD-SEOB} shows a
plot of $k/N$ versus $\hat{P}_k(\lambda_{k,\alpha})$ for all binary parameters.
For the top panel, both injections and inference are performed with the same model, and the recovered probability distribution is consistent with $P(<p)=p$, as expected.

\subsection{Zero noise runs to assess systematic biases}
\label{sec:sub:zeronoise}

Our synthetic data consists of expected detector responses $h(t)$ superimposed on detector noise realization $n(t)$.
The recovered posterior distribution's properties and in particular maximum-likelihood parameters depend on the specific
noise realization used.   To disentangle the deterministic effects of waveform systematics from the stochastic impact
of different noise realizations, we also repeat our analyses with the "zero noise'' realization: $n(t)=0$.

\subsection{Model-model mismatch}
\label{sec:sub:mismatch}

Several previous investigations (e.g.,
\cite{gr-nr-WaveformErrorStandards-LBO-2008,2009PhRvD..79l4033R,2010PhRvD..82h4020L,gwastro-mergers-HeeSuk-FisherMatrixWithAmplitudeCorrections,2010PhRvD..82l4052H,2016PhRvD..93j4050K,2020PhRvR...2b3151P}
and references therein) have phenomenologically argued that the magnitude of systematic
biases are related to the model-model
\emph{mismatch}, a simple inner-product-based estimate of waveform similarity between two model predictions
$h_1(\lambda)$ and $h_2(\lambda)$ at identical model parameters $\lambda$:
\begin{align}
{\cal M}(\lambda) = 1 - \max_{t_c,\phi_c} \frac{|\qmstateproduct{h_1}{e^{i(2\pi f t_c +\phi_c) }h_2}|}{|h_1||h_2|}
\end{align}
In this expression, the inner product
$\langle a|b\rangle_{k}\equiv
\int_{-\infty}^{\infty}2df\tilde{a}(f)^{*}\tilde{b}(f)/S_{h,k}(|f|)$ is  implied by the k$^{th}$ detector's
noise power spectrum $S_{h,k}(f)$, which for the purposes of waveform similarity is assumed to be the
advanced LIGO instrument, H1. In practice we adopt a low-frequency cutoff $f_{\rm min}$ so all inner products are modified to
\begin{equation}
\label{eq:overlap}
\langle a|b\rangle_{k}\equiv 2 \int_{|f|>f_{\rm min}}df\frac{[\tilde{a}(f)]^{*}\tilde{b}(f)}{S_{h,k}(|f|)}.
\end{equation}

\begin{figure}
\includegraphics[width=\columnwidth]{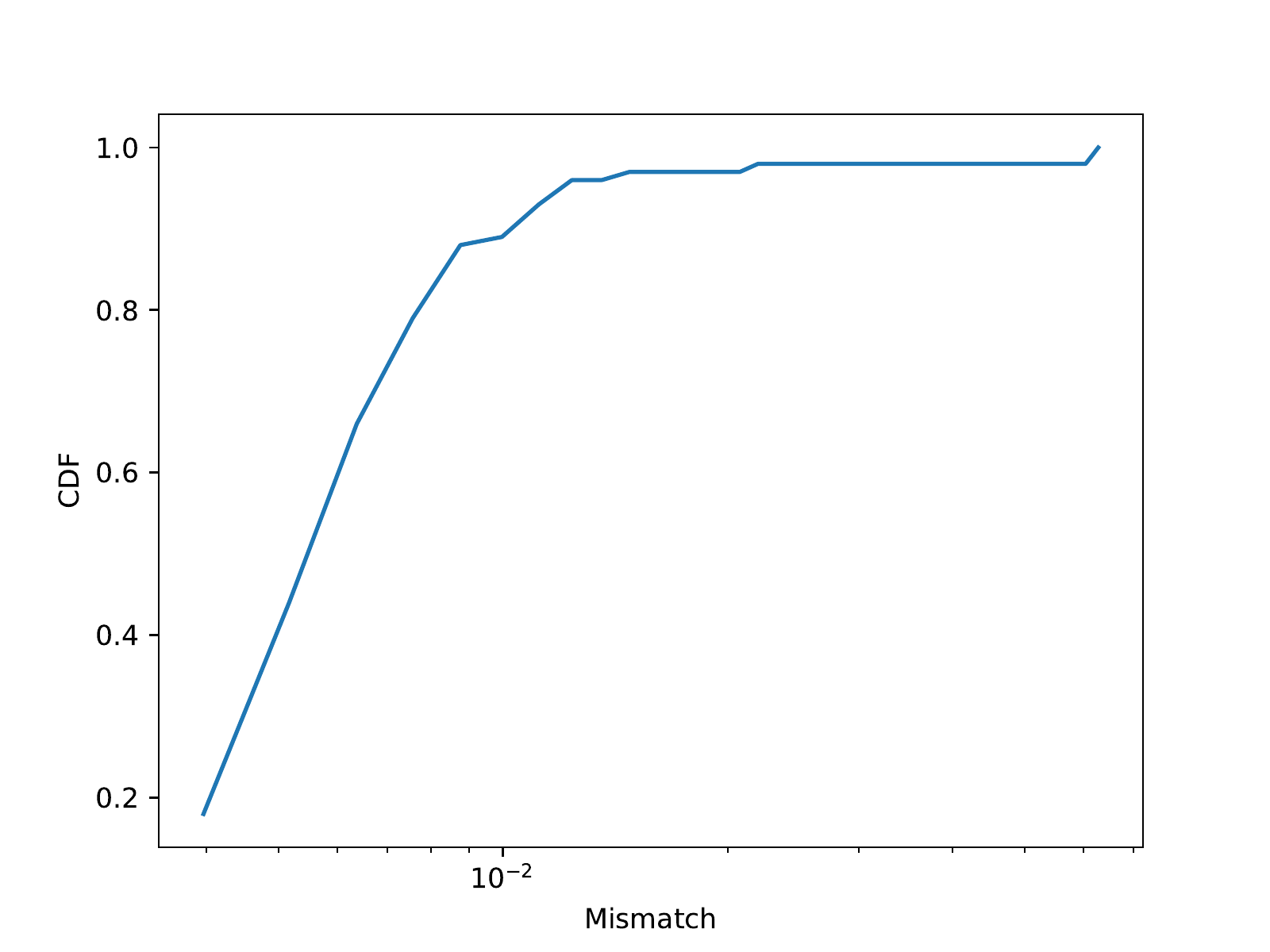}
\caption{\label{fig:Mismatch}\textbf{Cumulative Mismatch Distribution for population}: For all the synthetic sources in
  our population, we evaluate the GW strain along the $z$ axis using \SEOBA and  \IMRPD, then compute the mismatch
  between them.  This figure shows the cumulative distribution of these mismatches, most of which are slightly less than $10^{-2}$.}
\end{figure}

Figure  \ref{fig:Mismatch}  shows the distribution of mismatches for our synthetic population, where $h_1$ is generated
using \SEOBA and $h_2$ with \IMRPD.  For simplicity, we regenerate all signals at zero inclination, to
avoid polarization-related effects associated with the precise emission direction. For our fiducial compact binary population, 
the mismatches between these two
models are typically below $10^{-2}$, consistent with previous reports on systematic differences between these two
waveforms and with their similarity to even more accurate models and simulations \cite{gwastro-SEOBNRv4,2016PhRvD..93d4007K,Purrer:2019jcp}
% \editremark{citations, e.g. Bohe et al
%  arxiv:1611.03703 and Khan et al arxiv:1508.07253 (e.g., their Fig 19)}.

\subsection{Marginalizing over waveform systematics}
\label{sec:sub:MM}

Suppose we have two models $A$ and $B$ for GW strain, and use them to interpret a particular GW source.  We have prior
probabilities $p(A|\lambda)$ and $p(B|\lambda)$, characterizing our relative confidence in these two models for a source
with parameters $\lambda$.\footnote{For simplicity I will assume there are no internal model hyperparameters, but
  the method is easily generalized to include them.}   Suppose we have produced a RIFT analysis with each model for this
event, and have marginal likelihood functions ${\cal L}_{A}(\lambda)$ and ${\cal L}_B(\lambda)$ evaluated at a
\emph{single} point $\lambda$.  
We can therefore construct the marginal likelihood for $\lambda$ by averaging over both models:
\begin{align}
\label{eq:L:av}
{\cal L}_{av}(\lambda) = p(A|\lambda) {\cal L}_A(\lambda) + p(B|\lambda) {\cal L}_B(\lambda)
\end{align}
For simplicity the calculations in this work always adopt $p(A|\lambda)=p(B|\lambda)=1/2$.
We can therefore transparently integrate multi-model inference into RIFT as follows.  We assume we have a single grid
of points $\lambda_k$ such that both $(\lambda_k, {\cal L}_A(\lambda_k)$ and $(\lambda_k, {\cal L}_B(\lambda_k)$  can be
interpolated to produce reliable likelihoods and thus posterior distributions $p_A(\lambda)$ and $p_B(\lambda)$,
respectively.  At each point $\lambda_k$ we therefore construct ${\cal L}_{av}(\lambda_k)$ by the above procedure. 
We then interpolate to approximate $\hat{\cal L}(\lambda)$ versus the continuous parameters $\lambda$.

Operationally speaking, we construct model-averaged marginal likelihoods by the following procedure.   First, we
construct a fiducial grid for models A and B, for example by joining the grids used to independently analyze A and B.
We use ILE to evaluate ${\cal L}_A(\lambda_k)$ and ${\cal L}_B(\lambda_k)$ on this grid.  We construct ${\cal
  L}_{av}(\lambda_k)$ as above.  We use the combinations $(\lambda_k, {\cal L}_{av})$ with CIP, to construct a
model-averaged posterior distribution.

Our procedure bears considerable resemblance to the approach suggested by Ashton and Khan, but we have organized the
calculation differently.  In that approach, AK used   the evidences $Z_A = \int {\cal L}_A
p(\lambda) d\lambda $ and $Z_B$ for the two
waveform models.  While we can compute both quantities with very high accuracy, we prefer to directly average between
waveform models \emph{at the same choice of intrinsic parameters} (i.e., via Eq. (\ref{eq:L:av})) , to insure that marginalization over waveform models is completely decoupled from the interpolation
techniques used to construct $\hat{{\cal L}}$ from the sampled data.

\section{Results}
\label{sec:results}

Using our fiducial BBH population, we generated 100 synthetic signals using \IMRPD, and another 100 synthetic signals with \SEOBA.  
For each signal, we performed parameter inference with \emph{both} \IMRPD and \SEOBA. 
These inferences allow us both to assess the impact of waveform systematics in our fiducial population, and mitigate them.

\subsection{Demonstrating and quantifying waveform systematics}

\begin{figure}
\includegraphics[scale=0.4]{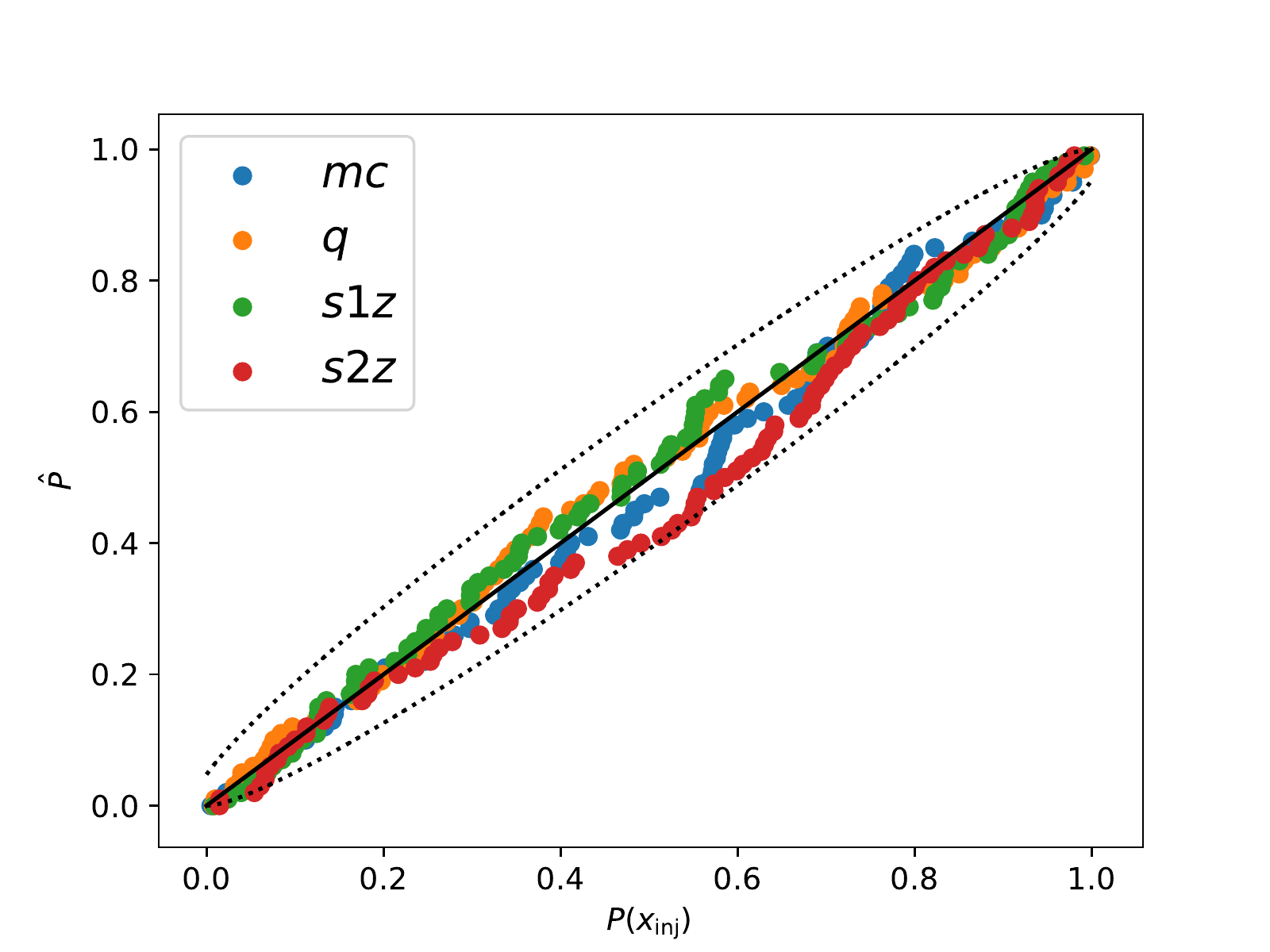}
\includegraphics[scale=0.4]{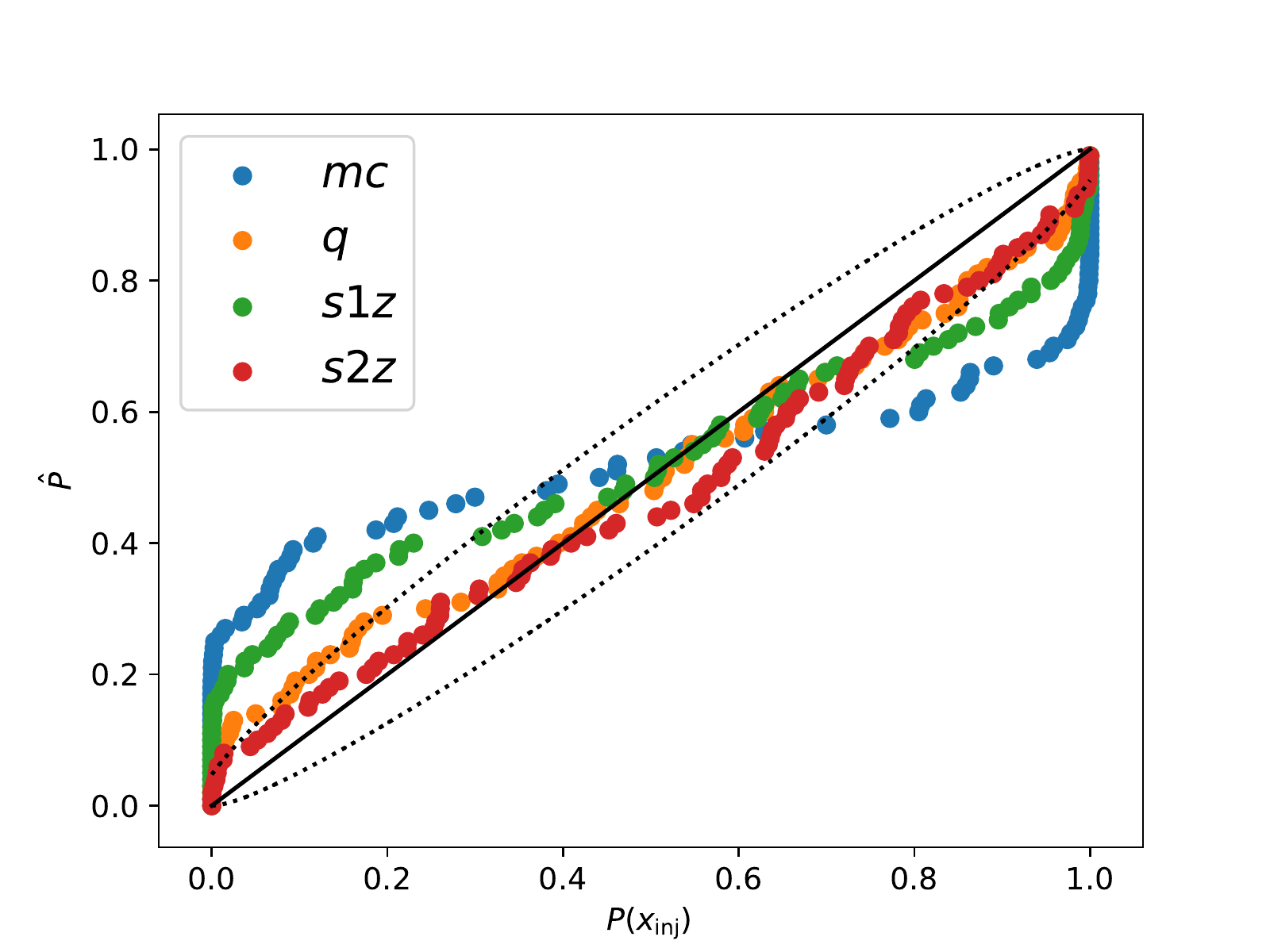}
\caption{PP-plot of events injected with \SEOBA and recovered with \SEOBA(top panel) and \IMRPD(bottom panel) waveform.
The dashed line indicates the 90\% credible interval expected for a cumulative distribution  drawn from 100
uniformly-distributed samples.
}
\label{pp-IMRD-SEOB}
\end{figure}

% PP plot diagnostic
The PP plot provides the most compelling demonstration of waveform systematics' pernicious impact.
Ideally, when recovering a known model and a known population, we expect to recover the injected values as often as they
occur, producing a diagonal PP plot.  The top panel of Figure \ref{pp-IMRD-SEOB} shows precisely what we expect, when we
inject and recover with the same model (here, \SEOBA).  
By contrast, the bottom panel shows a PP plot generated using inference from  \IMRPD on the same \SEOBA injections.
The PP plot is considerably non-diagonal, reflecting frequent and  substantial parameter biases in our fiducial population.

\begin{figure}[ht]
\includegraphics[width=\columnwidth]{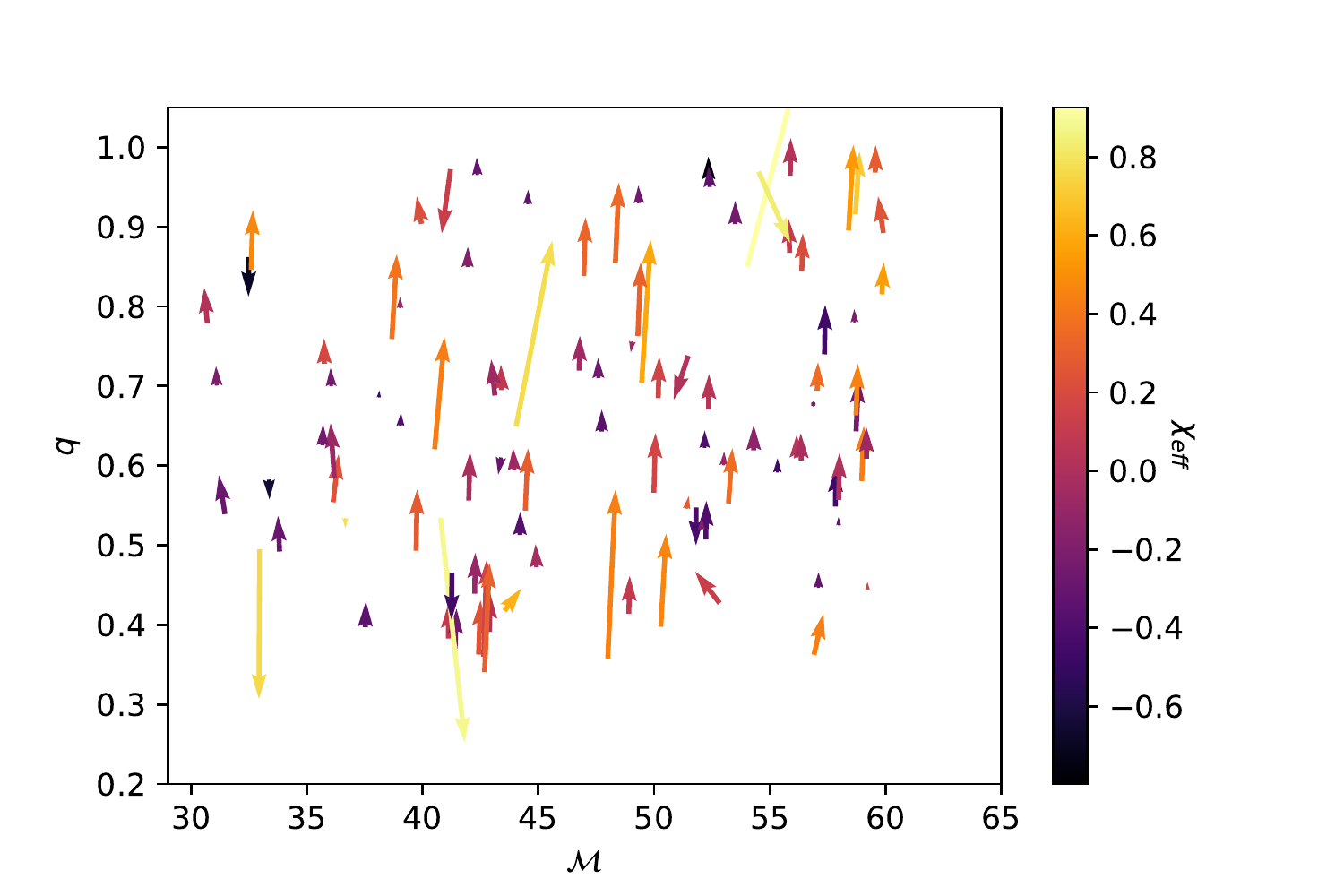}
\includegraphics[width=\columnwidth]{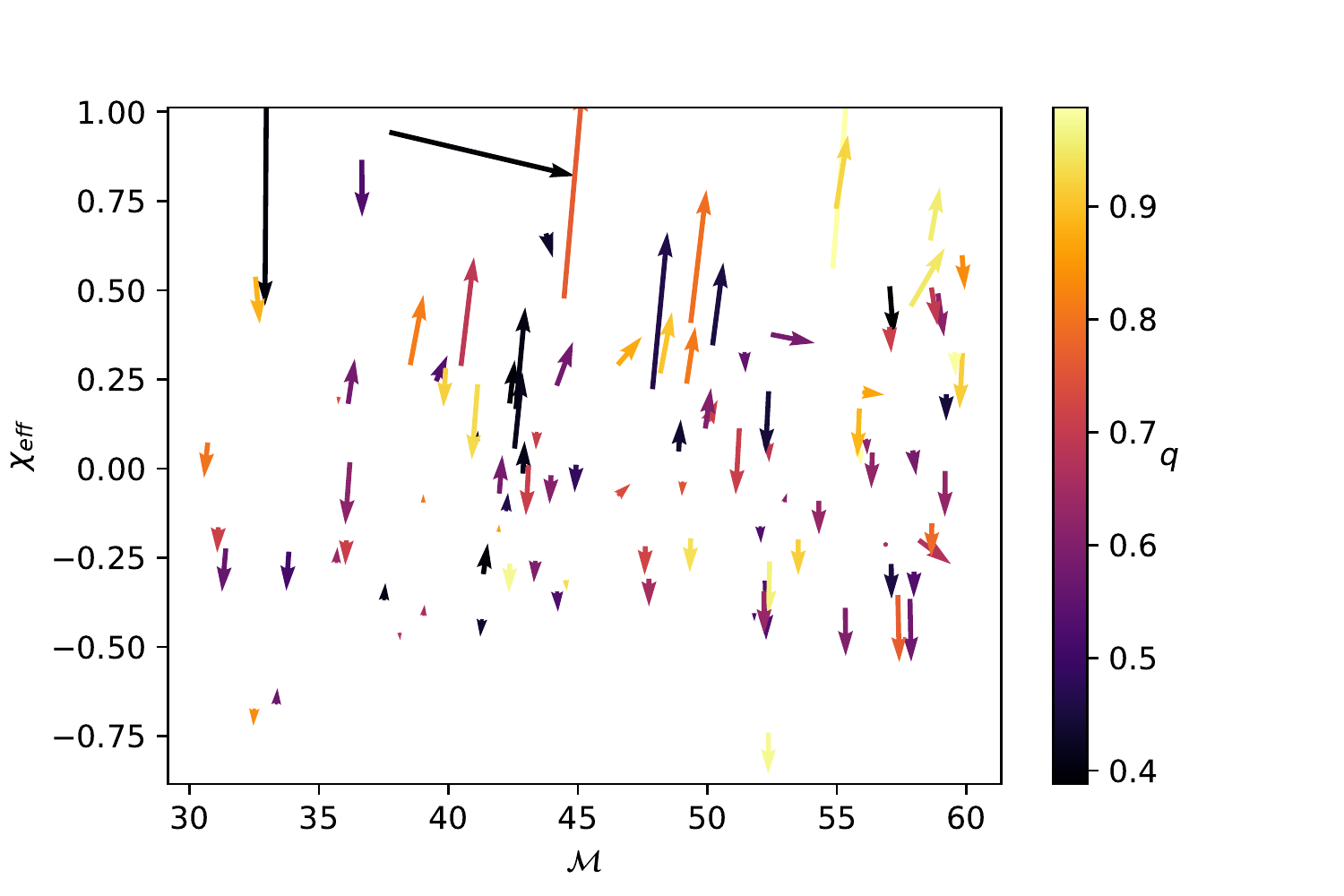}	
\includegraphics[width=\columnwidth]{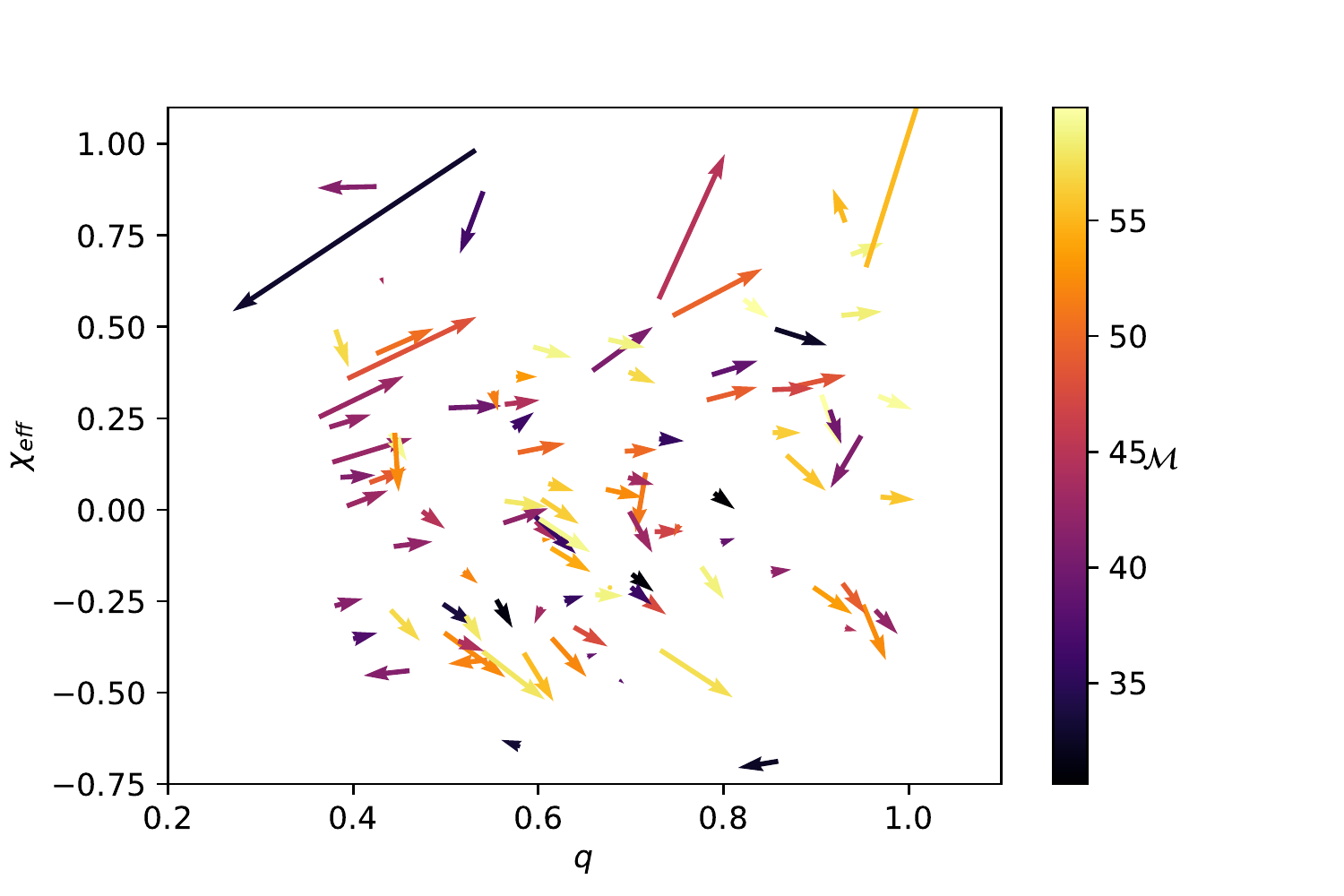}
\caption{Vector plot showing amplitude-scaled offsets between \SEOBA and \IMRPD for parameters $\mathcal{M}$ and $q$
  (top panel), $\mathcal{M}$ and $\chi_{\rm eff}$ (middle panel) and $q$ and $\chi_{\rm eff}$ (bottom panel) as a function of the respective parameters with color map being the value of the parameter mentioned on the color scale.}
\label{Quiver}
\end{figure}

Parameter biases introduced by waveform systematics vary in magnitude and direction over the parameter space.  To
illustrate these offsets for the parameters $x=\mc,q,\chi_{\rm eff}$, we've evaluated the parameter shift $\Delta x$
between the mean  inferred with \IMRPD{} and the mean inferred with \SEOBA,
\emph{relative} to $\sigma \rho$, which is a product of $\rho$ (the signal-to-noise ratio, a measure of the signal
amplitude) and  the statistical error (as measured by the standard
deviation $\sigma$ of the posterior of the parameter $x$ in question).  [The combination $\sigma \rho$ is approximately independent of signal amplitude, allowing
  us to measure the effect of waveform systematics for a fiducial amplitude.]
Figure \ref{Quiver} shows  a vector plot of these scaled offsets $\Delta x/\rho \sigma$, as a function of two of the parameters at a time.  
The length of the arrow corresponds to the scaled shifts in the parameters $\mathcal{M}$, $q$ and $\chi_{\rm eff}$, 
plotted against the injected parameter values.  The color scale shows the remaining parameter.  
The top two panels show that shifts in  $q=m_2/m_1$, $\chi_{\rm eff} = (m_1\chi_{1,z}+m_2\chi_{2,z})/(m_1+m_2)$ are substantial. Parameter shifts for $q$ generally increase
with $\chi_{\rm eff}$.  Shifts in $\chi_{\rm eff}$ are generally positive for positive $\chi_{\rm eff}$, negative for
negative $\chi_{\rm eff}$, and strongly dependent on mass ratio, with more substantial shifts at either comparable mass
or at very high mass ratio, respectively.  In both cases, chirp mass $\mc$ has modest impact, with somewhat larger shifts occurring
at somewhat larger values of chirp mass. 
Most extreme waveform systematics seem to be associated with large mass ratio.

\begin{figure}
\includegraphics[scale=0.4]{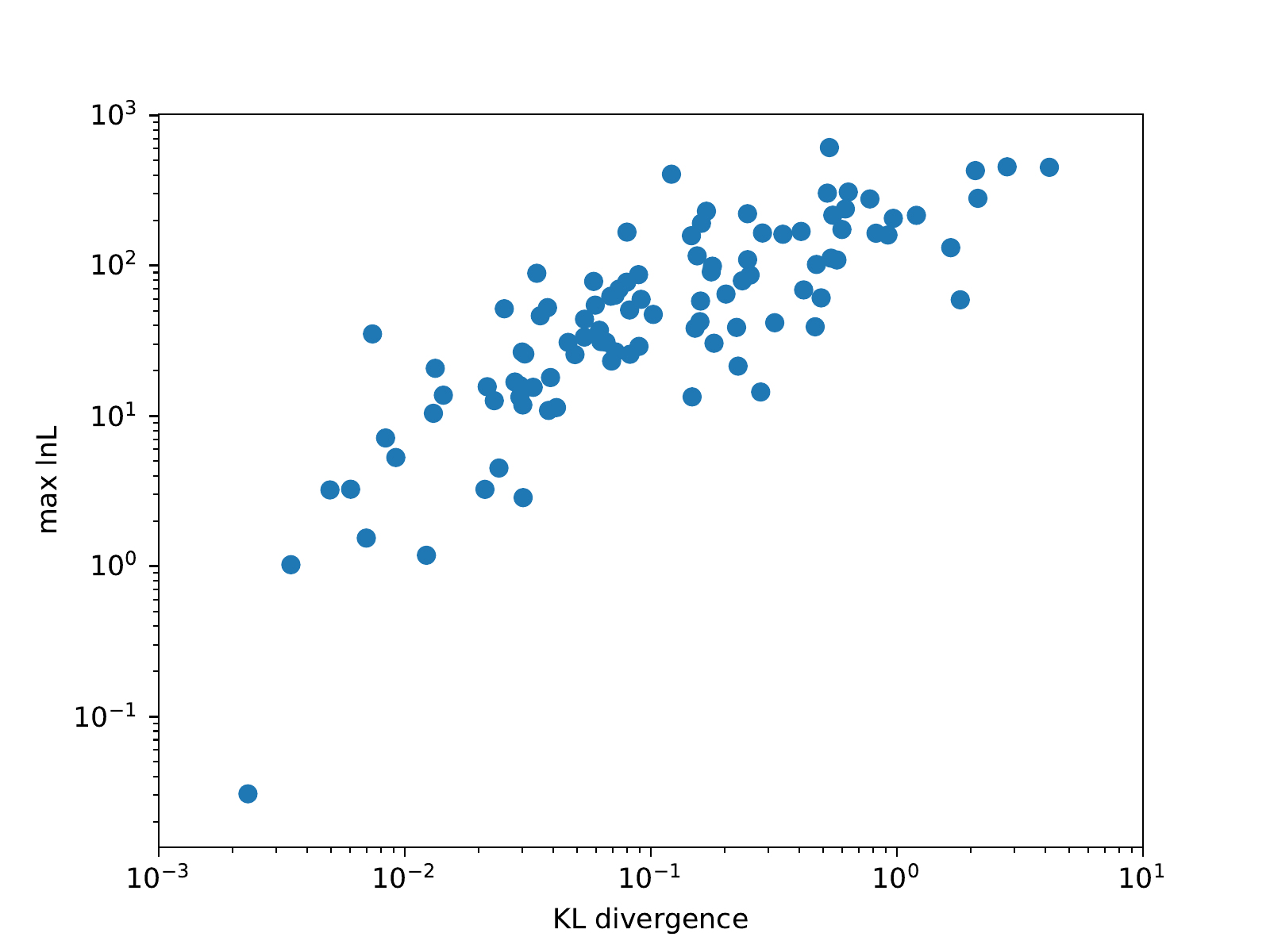}
\caption{Figure showing KL-divergences between the two waveform models versus the log of the maximum likelihood 
for the combined posteriors of $\mathcal{M}$, $\it{q}$ and $\chi_{\rm eff}$.}
\label{fig:Differences:KL}
\end{figure}

% Convergence diagnostic
Relative differences in mean value only imperfectly captures the differences between the two posteriors. As a sharper
diagnostic that includes parameter correlations, we use the mean and covariance of each distribution in $\mc,q,\chi_{\rm eff}$ to generate a local gaussian
approximation for each posterior, and then compute the KL divergence between these two gaussian approximations
%\editremark{cite RIFT paper, RIFT update}
\cite{gwastro-PENR-RIFT}. We expect more substantial differences and thus larger KL divergence for stronger
signals, whose posteriors are more sharply constrained.
To corroborate our intuition, Figure \ref{fig:Differences:KL} shows a scatterplot, with these KL divergences on the
horizontal axis and the largest value of $\ln {\cal L}$ on the vertical axis.  As expected, for the strongest signals,
differences between the two waveform models are the most pronounced.  %\editremark{further comments}. 

\begin{figure}
\includegraphics[scale=0.5]{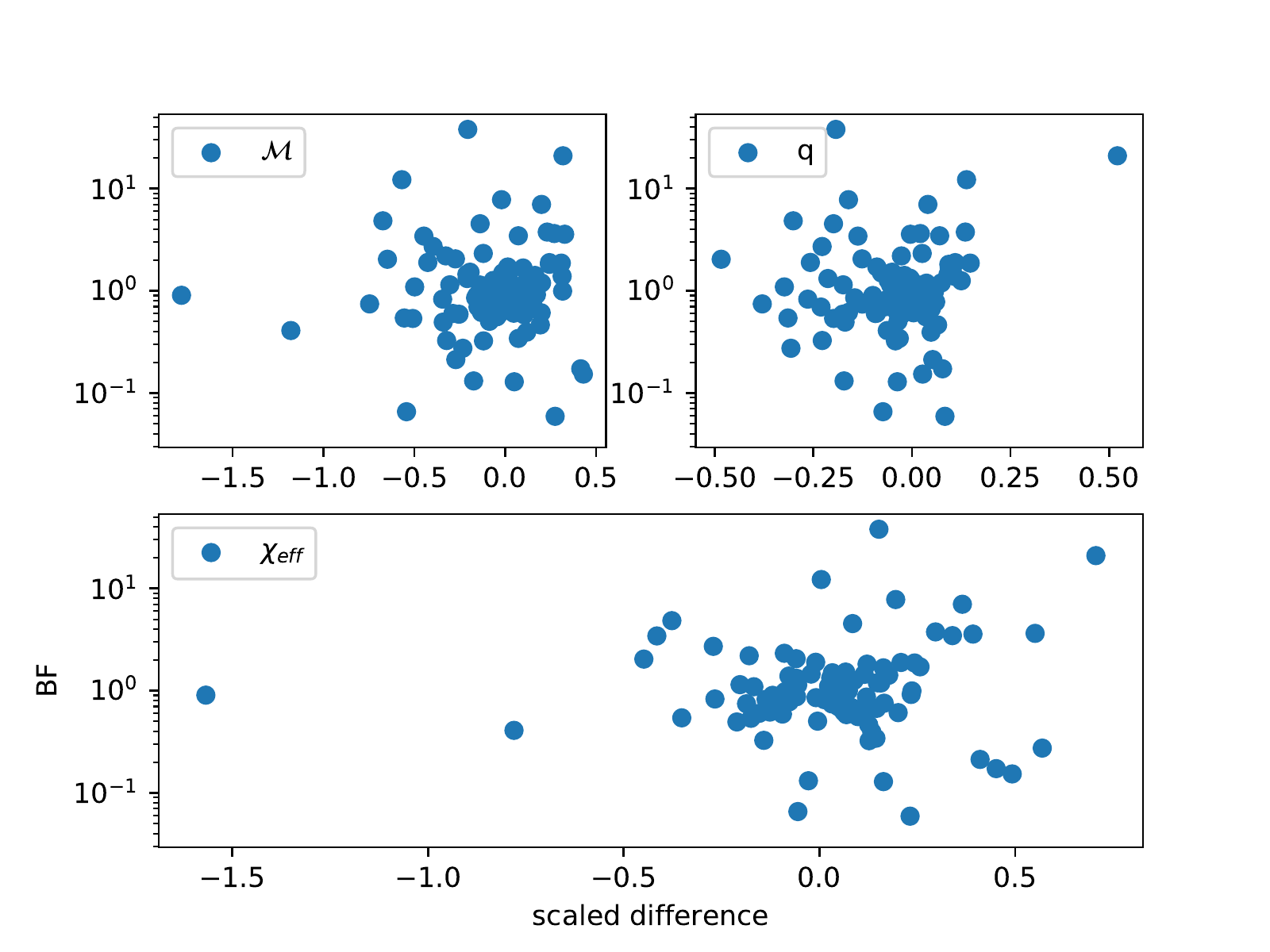}
\caption{Figure showing Bayes factor (BF) for \SEOBA versus \IMRPD plotted against differences between the \SEOBA and  \IMRPD waveforms for parameters $\mathcal{M}$, $\it{q}$ and $\chi_{\rm eff}$.}
\label{fig:Differences:Offsets}
\end{figure}

% Offsets results
One might expect that large parameter offsets are more likely to occur when the data favors one model or another.
While conceivably true asymptotically, for our specific synthetic population, we don't find a strong correlation between 
the Bayes factor ($\mathcal{Z}_{\SEOBA}/\mathcal{Z}_{\IMRPD}$) and any parameter offsets.  Figure \ref{fig:Differences:Offsets} shows this Bayes factor (BF) plotted versus the
scaled parameter offsets in $\mc,q,\chi_{\rm eff}$. Large offsets can occur without the data more strongly favoring
one model or the other, and vice versa.

\subsection{A PP plot test for marginalizing over waveform errors}
\begin{figure}
\includegraphics[scale=0.5]{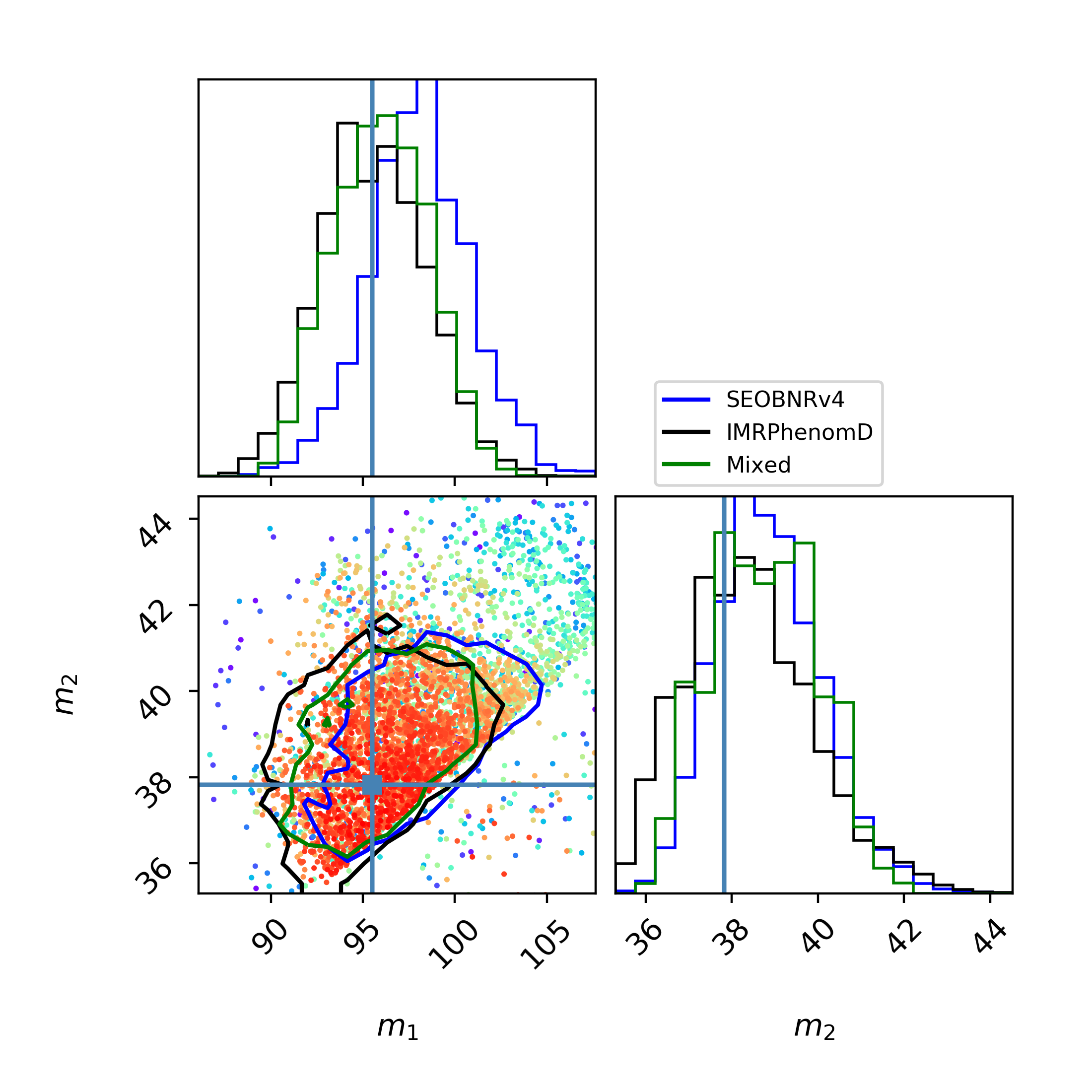}
\includegraphics[scale=0.45]{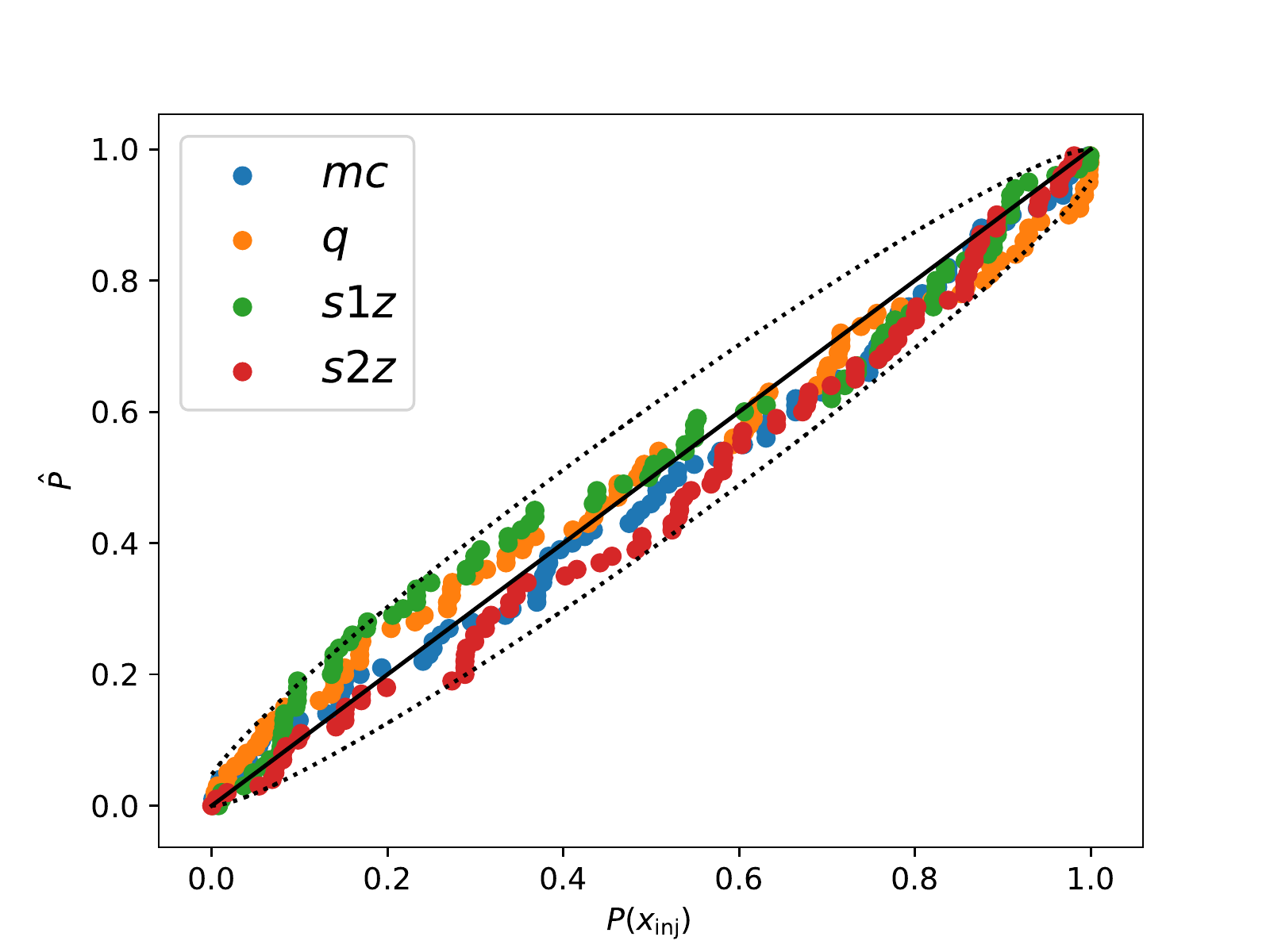}
\caption{\label{fig:ModelAveraging:PP} Top panel: example of a model averaged result.  The blue and black curves show
  the 1D marginal distributions and 2D 90\% credibles for \SEOBA and \IMRPD inferences, respectively. The green curves
  show the corresponding model-averaged result.  Bottom panel: PP plot test for our model-marginalized procedure.}
\end{figure}

% https://dcc.ligo.org/LIGO-G2001500
We test our model-averaged waveform procedure using a full synthetic PP plot procedure.  Specifically, we use the  $n_s=100$
synthetic source parameters. For each source, we pick one waveform model $A,B$ with probabilities $p(A),p(B)$, and  use it to
generate the signal. We then analyze the signal using the model-averaged procedure described above.  
As a concrete example, the top panel of Figure \ref{fig:ModelAveraging:PP} shows our analysis of one fiducial event in our synthetic sample.
The colored points show likelihood evaluations, with color scale corresponding to the marginalized likelihood evaluated with
\IMRPD.  The blue and black contours show the 90\% credible intervals for \SEOBA and
 \IMRPD, respectively; the two posteriors differ substantially, illustrating the impact of model systematics on
parameter inference.   The green contour shows our model-marginalized posterior.
  %% \editremark{shouldn't it be an average
  %% of the two posteriors? It won't exactly be a weighted average, because the grid is different, but it should be close
  %% ...}. 
 For comparison, the cross shows
the injected source parameters, and the model was \IMRPD. 
%% In the corner plot, there is a clear difference between the blue and black(expected) contours,
%% and green is the model marginalized posterior and it clearly similar to black->improve/remove. 

The bottom panel of Figure  \ref{fig:ModelAveraging:PP} shows one PP plot corresponding to applying our model-marginalized
procedure to a population where each source is randomly selected from either  \IMRPD or \SEOBA.  The dotted line shows a
90\% frequentist interval for the largest of four random cumulative distributions.  This figure shows our PP plots are
consistent with the diagonal, as desired.

\section{Discussion}
 
 In this work, we performed simple tests which
 reproduce significant differences between the models \SEOBA and \IMRPD, and can be extended to other available waveforms easily using 
 RIFT, an efficient parameter estimation engine. The probability-probability (PP) plot test, a commonly used statistical test, can be used to
 confirm differences between waveform models and as shown in Fig. \ref{pp-IMRD-SEOB}, parameter estimation performed using a 
 model different from  the injected model, gives a non-diagonal pp-plot for most parameters. We calculated the magnitude and direction of the offsets 
 introduced due to using a waveform model different to the injected model, and these differences are higher for extreme case scenarios, as expected. 
 A linear correlation between the KL divergence computed for the two models and the log of the maximum likelihood of the injected model,
 shows that high-SNR signal will have larger differences in the inferred parameter from various models. 
Because the most informative signals exhibit the largest parameter biases, waveform systematics have the potential to
strongly contaminate population inference.
Most importantly, we also demonstrated a method to mitigate these waveform systematics by marginalizing over the models used for parameter estimation
 analyses. 
 
 %As future detectors are expected to become 
 %more sensitive, the chances of high-SNR events will be high, these systematic differences between models bring in biases, which in turn 
 %will bring in biases in inference of the population properties of the compact objects

Our method requires as input some prior probabilities $p(X_k|\lambda)$ for different waveform models $X_k$.  One way
these  prior
probabilities could be selected is by  waveform faithfulness studies between models and numerical
relativity simulations.  These fidelity studies inevitably suggest waveform models vary in reliability over their
parameter space (e.g. \cite{2014PhRvD..89d2002K,2015PhRvD..92j2001K}), suggesting $p(X_k|\lambda)$ will depend nontrivially on $\lambda$.  
Operationally, these model priors propagate into each model's posterior inferences as if parameter inferences for model $X$ are performed using a
model-dependent prior $\propto p_{\rm prior}(\lambda)p(X_k|\lambda)$, instead of a common prior for all models. 
RIFT can seamlessly perform these calculations at minimal added computational expense, while simultaneously returning
results for each model derived from the conventional prior alone.

\section{Conclusions}
\label{sec:conclude}

%% In this paper, we propose to study the effects different waveform models have on the parameter estimation 
%% of the gravitational wave signal by utilizing tools like probability-probability plots(PP-plots), model averaging
%% and mismatch calculation. We will generate a large, random suite of synthetic signals, performing parameter
%%  inference with a variety of approximations. We will produce single-event and
%% population-based diagnostics for the systematic biases in parameter recovery 
%%  introduced by these models. We will focus on differences between detailed time-domain
%% models and fast but less comprehensive approximations.
 
Many waveform models exist currently that describe compact binary coalescences. Even though these are 
 derived by solving Einstein's equations, the various analytical or numerical approximation considered bring in differences and affect the parameter
 estimation process leading to biased interpretation of results. Averaging over the waveform models can mitigate these biases.
% that arise precisely due to the various differences in the waveform models.  
Building on prior directly comparable work \cite{Ashton:2019leq}, we have demonstrated an efficient method to perform such model marginalization.

Other techniques have been proposed to marginalize over waveform model systematics.  Notably, several groups have
proposed using the error estimates provided by their model regressions (e.g., the gaussian process error)
\cite{2020PhRvD.101d4027C}.   Relative to regression-based methods, our method has two notable advantages.
Our method can be immediately generalized to include  multiple  waveform models.  Critically, we  plan to introduce
parameter-dependent weighting of the likelihood from a waveform, since different waveforms are accurate in different
regimes.   No other model-marginalization technique can presently provide this level of control.

%% We also plan to extend the model-marginalization method to demonstrate and mitigate differences between \SEOBA and 
%% it's surrogate frequency-domain version SEOBNRv4\_ROM \cite{gwastro-SEOBNRv4}, which has been studied 
%% to have significant waveform systematics \editremark{citation} and also implement multiple-model marginalization. 
%% We also aim to get analytical approximation(s) for the vector offsets shown in Fig. \ref{fig:Differences:Offsets}, 
%% which can be used to correct for biases introduced when using either of the models. Extension to BNS regime as well?
 
%% The concept of marginalizing over waveform models has also been demonstrated by Ashton and Khan\cite{Ashton:2019leq}
%% as described in the \ref{sec:Intro} and by Chua et al\cite{2020PhRvD.101d4027C}. \editremark{add more} 

%Other methods for marginalizing, by directly allowing for uncertainty in the waveform model (e.g., GP-based methods)
%* Chua et al 1912.11543  \cite{2020PhRvD.101d4027C}

% CONTEXT PAPERS
%* Ashton and Khan arxiv:1910.09138 \cite{Ashton:2019leq}  -- they use the same marginalization idea

%% * Puerrer and Haster \cite{Purrer:2019jcp} (``ready for what lies ahead'' paper)
%% \href{https://journals.aps.org/prresearch/abstract/10.1103/PhysRevResearch.2.023151}{link}: simple analytic estimates
%% for the most part, with a few PE examples (i.e, biases are huge), and Fig 5 and 6  (see Eq. 11,just Pv2)
%** significant biases in O3-scale IFOs per event, but they did not find huge bias examples
%** discussed mismatch /residual power

\begin{acknowledgements}
ROS  gratefully acknowledges support from NSF awards PHY-1707965, PHY-2012057, and AST-1909534.
AY acknowledges support from NSF PHY-2012057 grant. The authors are grateful for computational resources provided by 
the LIGO Laboratories at CIT and LHO supported by National Science
Foundation Grants PHY0757058 and PHY-0823459.
%\editremark{get boilerplate text}
\end{acknowledgements}

%\appendix
%\section{A procedure for marginalizing over waveform errors}

%\bibstyle{unsrt}
\bibliography{references}

\begin{thebibliography}{44}
\expandafter\ifx\csname natexlab\endcsname\relax\def\natexlab#1{#1}\fi
\expandafter\ifx\csname bibnamefont\endcsname\relax
  \def\bibnamefont#1{#1}\fi
\expandafter\ifx\csname bibfnamefont\endcsname\relax
  \def\bibfnamefont#1{#1}\fi
\expandafter\ifx\csname citenamefont\endcsname\relax
  \def\citenamefont#1{#1}\fi
\expandafter\ifx\csname url\endcsname\relax
  \def\url#1{\texttt{#1}}\fi
\expandafter\ifx\csname urlprefix\endcsname\relax\def\urlprefix{URL }\fi
\providecommand{\bibinfo}[2]{#2}
\providecommand{\eprint}[2][]{\url{#2}}

\bibitem[{\citenamefont{{Abbott et al. (The LIGO Scientific Collaboration and
  the Virgo Collaboration)}}(2016)}]{DiscoveryPaper}
\bibinfo{author}{\bibfnamefont{B.}~\bibnamefont{{Abbott et al. (The LIGO
  Scientific Collaboration and the Virgo Collaboration)}}},
  \bibinfo{journal}{\prl} \textbf{\bibinfo{volume}{16}},
  \bibinfo{pages}{061102} (\bibinfo{year}{2016}).

\bibitem[{\citenamefont{{LIGO Scientific Collaboration}
  et~al.}(2015)\citenamefont{{LIGO Scientific Collaboration}, {Aasi}, {Abbott},
  {Abbott}, {Abbott}, {Abernathy}, {Ackley}, {Adams}, {Adams}, {Addesso}
  et~al.}}]{2015CQGra..32g4001L}
\bibinfo{author}{\bibnamefont{{LIGO Scientific Collaboration}}},
  \bibinfo{author}{\bibfnamefont{J.}~\bibnamefont{{Aasi}}},
  \bibinfo{author}{\bibfnamefont{B.~P.} \bibnamefont{{Abbott}}},
  \bibinfo{author}{\bibfnamefont{R.}~\bibnamefont{{Abbott}}},
  \bibinfo{author}{\bibfnamefont{T.}~\bibnamefont{{Abbott}}},
  \bibinfo{author}{\bibfnamefont{M.~R.} \bibnamefont{{Abernathy}}},
  \bibinfo{author}{\bibfnamefont{K.}~\bibnamefont{{Ackley}}},
  \bibinfo{author}{\bibfnamefont{C.}~\bibnamefont{{Adams}}},
  \bibinfo{author}{\bibfnamefont{T.}~\bibnamefont{{Adams}}},
  \bibinfo{author}{\bibfnamefont{P.}~\bibnamefont{{Addesso}}},
  \bibnamefont{et~al.}, \bibinfo{journal}{Classical and Quantum Gravity}
  \textbf{\bibinfo{volume}{32}}, \bibinfo{eid}{074001} (\bibinfo{year}{2015}),
  \eprint{1411.4547}.

\bibitem[{\citenamefont{{Accadia} and {et
  al}}(2012)}]{gw-detectors-Virgo-original-preferred}
\bibinfo{author}{\bibfnamefont{T.}~\bibnamefont{{Accadia}}} \bibnamefont{and}
  \bibinfo{author}{\bibnamefont{{et al}}}, \bibinfo{journal}{Journal of
  Instrumentation} \textbf{\bibinfo{volume}{7}}, \bibinfo{pages}{P03012}
  (\bibinfo{year}{2012}),
  \urlprefix\url{http://iopscience.iop.org/1748-0221/7/03/P03012}.

\bibitem[{\citenamefont{Acernese et~al.}(2015)}]{TheVirgo:2014hva}
\bibinfo{author}{\bibfnamefont{F.}~\bibnamefont{Acernese}} \bibnamefont{et~al.}
  (\bibinfo{collaboration}{VIRGO}), \bibinfo{journal}{Class. Quant. Grav.}
  \textbf{\bibinfo{volume}{32}}, \bibinfo{pages}{024001}
  (\bibinfo{year}{2015}), \eprint{1408.3978}.

\bibitem[{\citenamefont{{The LIGO Scientific Collaboration}
  et~al.}(2020{\natexlab{a}})\citenamefont{{The LIGO Scientific Collaboration},
  {the Virgo Collaboration}, {Abbott}, {Abbott}, {Abbott}, {Abraham},
  {Acernese}, {Ackley}, {Adams}, {Adya} et~al.}}]{LIGO-O3-GW190521-discovery}
\bibinfo{author}{\bibnamefont{{The LIGO Scientific Collaboration}}},
  \bibinfo{author}{\bibnamefont{{the Virgo Collaboration}}},
  \bibinfo{author}{\bibfnamefont{B.~P.} \bibnamefont{{Abbott}}},
  \bibinfo{author}{\bibfnamefont{R.}~\bibnamefont{{Abbott}}},
  \bibinfo{author}{\bibfnamefont{T.~D.} \bibnamefont{{Abbott}}},
  \bibinfo{author}{\bibfnamefont{S.}~\bibnamefont{{Abraham}}},
  \bibinfo{author}{\bibfnamefont{F.}~\bibnamefont{{Acernese}}},
  \bibinfo{author}{\bibfnamefont{K.}~\bibnamefont{{Ackley}}},
  \bibinfo{author}{\bibfnamefont{C.}~\bibnamefont{{Adams}}},
  \bibinfo{author}{\bibfnamefont{V.~B.} \bibnamefont{{Adya}}},
  \bibnamefont{et~al.}, \bibinfo{journal}{\prl} \textbf{\bibinfo{volume}{125}},
  \bibinfo{eid}{101102} (\bibinfo{year}{2020}{\natexlab{a}}).

\bibitem[{\citenamefont{{The LIGO Scientific Collaboration}
  et~al.}(2020{\natexlab{b}})\citenamefont{{The LIGO Scientific Collaboration},
  {the Virgo Collaboration}, {Abbott}, {Abbott}, {Abbott}, {Abraham},
  {Acernese}, {Ackley}, {Adams}, {Adya}
  et~al.}}]{LIGO-O3-GW190521-implications}
\bibinfo{author}{\bibnamefont{{The LIGO Scientific Collaboration}}},
  \bibinfo{author}{\bibnamefont{{the Virgo Collaboration}}},
  \bibinfo{author}{\bibfnamefont{B.~P.} \bibnamefont{{Abbott}}},
  \bibinfo{author}{\bibfnamefont{R.}~\bibnamefont{{Abbott}}},
  \bibinfo{author}{\bibfnamefont{T.~D.} \bibnamefont{{Abbott}}},
  \bibinfo{author}{\bibfnamefont{S.}~\bibnamefont{{Abraham}}},
  \bibinfo{author}{\bibfnamefont{F.}~\bibnamefont{{Acernese}}},
  \bibinfo{author}{\bibfnamefont{K.}~\bibnamefont{{Ackley}}},
  \bibinfo{author}{\bibfnamefont{C.}~\bibnamefont{{Adams}}},
  \bibinfo{author}{\bibfnamefont{V.~B.} \bibnamefont{{Adya}}},
  \bibnamefont{et~al.}, \bibinfo{journal}{arXiv e-prints}
  \bibinfo{eid}{arXiv:2009.01190} (\bibinfo{year}{2020}{\natexlab{b}}),
  \eprint{2009.01190}.

\bibitem[{\citenamefont{{The LIGO Scientific Collaboration}
  et~al.}(2020{\natexlab{c}})\citenamefont{{The LIGO Scientific Collaboration},
  {the Virgo Collaboration}, {Abbott}, {Abbott}, {Abbott}, {Abraham},
  {Acernese}, {Ackley}, {Adams}, {Adya} et~al.}}]{LIGO-O3-GW190814}
\bibinfo{author}{\bibnamefont{{The LIGO Scientific Collaboration}}},
  \bibinfo{author}{\bibnamefont{{the Virgo Collaboration}}},
  \bibinfo{author}{\bibfnamefont{B.~P.} \bibnamefont{{Abbott}}},
  \bibinfo{author}{\bibfnamefont{R.}~\bibnamefont{{Abbott}}},
  \bibinfo{author}{\bibfnamefont{T.~D.} \bibnamefont{{Abbott}}},
  \bibinfo{author}{\bibfnamefont{S.}~\bibnamefont{{Abraham}}},
  \bibinfo{author}{\bibfnamefont{F.}~\bibnamefont{{Acernese}}},
  \bibinfo{author}{\bibfnamefont{K.}~\bibnamefont{{Ackley}}},
  \bibinfo{author}{\bibfnamefont{C.}~\bibnamefont{{Adams}}},
  \bibinfo{author}{\bibfnamefont{V.~B.} \bibnamefont{{Adya}}},
  \bibnamefont{et~al.}, \bibinfo{journal}{\apjl}
  \textbf{\bibinfo{volume}{896}}, \bibinfo{pages}{L44}
  (\bibinfo{year}{2020}{\natexlab{c}}),
  \urlprefix\url{https://doi.org/10.3847%2F2041-8213%2Fab960f}.

\bibitem[{\citenamefont{{The LIGO Scientific Collaboration}
  et~al.}(2020{\natexlab{d}})\citenamefont{{The LIGO Scientific Collaboration},
  {the Virgo Collaboration}, {Abbott}, {Abbott}, {Abbott}, {Abraham},
  {Acernese}, {Ackley}, {Adams}, {Adya} et~al.}}]{LIGO-O3-GW190412}
\bibinfo{author}{\bibnamefont{{The LIGO Scientific Collaboration}}},
  \bibinfo{author}{\bibnamefont{{the Virgo Collaboration}}},
  \bibinfo{author}{\bibfnamefont{B.~P.} \bibnamefont{{Abbott}}},
  \bibinfo{author}{\bibfnamefont{R.}~\bibnamefont{{Abbott}}},
  \bibinfo{author}{\bibfnamefont{T.~D.} \bibnamefont{{Abbott}}},
  \bibinfo{author}{\bibfnamefont{S.}~\bibnamefont{{Abraham}}},
  \bibinfo{author}{\bibfnamefont{F.}~\bibnamefont{{Acernese}}},
  \bibinfo{author}{\bibfnamefont{K.}~\bibnamefont{{Ackley}}},
  \bibinfo{author}{\bibfnamefont{C.}~\bibnamefont{{Adams}}},
  \bibinfo{author}{\bibfnamefont{V.~B.} \bibnamefont{{Adya}}},
  \bibnamefont{et~al.}, \bibinfo{journal}{\prd} \textbf{\bibinfo{volume}{102}},
  \bibinfo{eid}{043015} (\bibinfo{year}{2020}{\natexlab{d}}).

\bibitem[{\citenamefont{{The LIGO Scientific Collaboration}
  et~al.}(2020{\natexlab{e}})\citenamefont{{The LIGO Scientific Collaboration},
  {the Virgo Collaboration}, {Abbott}, {Abbott}, {Abbott}, {Abraham},
  {Acernese}, {Ackley}, {Adams}, {Adya} et~al.}}]{LIGO-O3-O3a-catalog}
\bibinfo{author}{\bibnamefont{{The LIGO Scientific Collaboration}}},
  \bibinfo{author}{\bibnamefont{{the Virgo Collaboration}}},
  \bibinfo{author}{\bibfnamefont{B.~P.} \bibnamefont{{Abbott}}},
  \bibinfo{author}{\bibfnamefont{R.}~\bibnamefont{{Abbott}}},
  \bibinfo{author}{\bibfnamefont{T.~D.} \bibnamefont{{Abbott}}},
  \bibinfo{author}{\bibfnamefont{S.}~\bibnamefont{{Abraham}}},
  \bibinfo{author}{\bibfnamefont{F.}~\bibnamefont{{Acernese}}},
  \bibinfo{author}{\bibfnamefont{K.}~\bibnamefont{{Ackley}}},
  \bibinfo{author}{\bibfnamefont{C.}~\bibnamefont{{Adams}}},
  \bibinfo{author}{\bibfnamefont{V.~B.} \bibnamefont{{Adya}}},
  \bibnamefont{et~al.}, \bibinfo{journal}{Available as LIGO-P2000061}
  (\bibinfo{year}{2020}{\natexlab{e}}),
  \urlprefix\url{https://dcc.ligo.org/LIGO-P2000061}.

\bibitem[{\citenamefont{Shaik et~al.}(2019)\citenamefont{Shaik, Lange, Field,
  O'Shaughnessy, Varma, Kidder, Pfeiffer, and Wysocki}}]{Shaik:2019dym}
\bibinfo{author}{\bibfnamefont{F.~H.} \bibnamefont{Shaik}},
  \bibinfo{author}{\bibfnamefont{J.}~\bibnamefont{Lange}},
  \bibinfo{author}{\bibfnamefont{S.~E.} \bibnamefont{Field}},
  \bibinfo{author}{\bibfnamefont{R.}~\bibnamefont{O'Shaughnessy}},
  \bibinfo{author}{\bibfnamefont{V.}~\bibnamefont{Varma}},
  \bibinfo{author}{\bibfnamefont{L.~E.} \bibnamefont{Kidder}},
  \bibinfo{author}{\bibfnamefont{H.~P.} \bibnamefont{Pfeiffer}},
  \bibnamefont{and} \bibinfo{author}{\bibfnamefont{D.}~\bibnamefont{Wysocki}}
  (\bibinfo{year}{2019}), \eprint{1911.02693}.

\bibitem[{\citenamefont{{Williamson} et~al.}(2017)\citenamefont{{Williamson},
  {Lange}, {O'Shaughnessy}, {Clark}, {Kumar}, {Bustillo}, and
  {Veitch}}}]{gwastro-Systematics-Williamson2017}
\bibinfo{author}{\bibfnamefont{A.}~\bibnamefont{{Williamson}}},
  \bibinfo{author}{\bibfnamefont{J.}~\bibnamefont{{Lange}}},
  \bibinfo{author}{\bibfnamefont{R.}~\bibnamefont{{O'Shaughnessy}}},
  \bibinfo{author}{\bibfnamefont{J.}~\bibnamefont{{Clark}}},
  \bibinfo{author}{\bibfnamefont{P.}~\bibnamefont{{Kumar}}},
  \bibinfo{author}{\bibfnamefont{J.}~\bibnamefont{{Bustillo}}},
  \bibnamefont{and} \bibinfo{author}{\bibfnamefont{J.}~\bibnamefont{{Veitch}}},
  \bibinfo{journal}{\prd} \textbf{\bibinfo{volume}{96}},
  \bibinfo{pages}{124041} (\bibinfo{year}{2017}),
  \urlprefix\url{https://journals.aps.org/prd/abstract/10.1103/PhysRevD.96.124041}.

\bibitem[{\citenamefont{Pürrer and Haster}(2020)}]{Purrer:2019jcp}
\bibinfo{author}{\bibfnamefont{M.}~\bibnamefont{Pürrer}} \bibnamefont{and}
  \bibinfo{author}{\bibfnamefont{C.-J.} \bibnamefont{Haster}},
  \bibinfo{journal}{Phys. Rev. Res.} \textbf{\bibinfo{volume}{2}},
  \bibinfo{pages}{023151} (\bibinfo{year}{2020}), \eprint{1912.10055}.

\bibitem[{\citenamefont{{Hannam} et~al.}(2014)\citenamefont{{Hannam},
  {Schmidt}, {Boh{\'e}}, {Haegel}, {Husa}, {Ohme}, {Pratten}, and
  {P{\"u}rrer}}}]{gwastro-mergers-IMRPhenomP}
\bibinfo{author}{\bibfnamefont{M.}~\bibnamefont{{Hannam}}},
  \bibinfo{author}{\bibfnamefont{P.}~\bibnamefont{{Schmidt}}},
  \bibinfo{author}{\bibfnamefont{A.}~\bibnamefont{{Boh{\'e}}}},
  \bibinfo{author}{\bibfnamefont{L.}~\bibnamefont{{Haegel}}},
  \bibinfo{author}{\bibfnamefont{S.}~\bibnamefont{{Husa}}},
  \bibinfo{author}{\bibfnamefont{F.}~\bibnamefont{{Ohme}}},
  \bibinfo{author}{\bibfnamefont{G.}~\bibnamefont{{Pratten}}},
  \bibnamefont{and}
  \bibinfo{author}{\bibfnamefont{M.}~\bibnamefont{{P{\"u}rrer}}},
  \bibinfo{journal}{\prl} \textbf{\bibinfo{volume}{113}}, \bibinfo{eid}{151101}
  (\bibinfo{year}{2014}), \eprint{1308.3271}.

\bibitem[{\citenamefont{{Khan} et~al.}(2019)\citenamefont{{Khan},
  {Chatziioannou}, {Hannam}, and {Ohme}}}]{gwastro-mergers-IMRPhenomPv3}
\bibinfo{author}{\bibfnamefont{S.}~\bibnamefont{{Khan}}},
  \bibinfo{author}{\bibfnamefont{K.}~\bibnamefont{{Chatziioannou}}},
  \bibinfo{author}{\bibfnamefont{M.}~\bibnamefont{{Hannam}}}, \bibnamefont{and}
  \bibinfo{author}{\bibfnamefont{F.}~\bibnamefont{{Ohme}}},
  \bibinfo{journal}{\prd} \textbf{\bibinfo{volume}{100}}, \bibinfo{eid}{024059}
  (\bibinfo{year}{2019}), \eprint{1809.10113}.

\bibitem[{\citenamefont{{Boh{\'e}} et~al.}(2017)\citenamefont{{Boh{\'e}},
  {Shao}, {Taracchini}, {Buonanno}, {Babak}, {Harry}, {Hinder}, {Ossokine},
  {P{\"u}rrer}, {Raymond} et~al.}}]{gwastro-SEOBNRv4}
\bibinfo{author}{\bibfnamefont{A.}~\bibnamefont{{Boh{\'e}}}},
  \bibinfo{author}{\bibfnamefont{L.}~\bibnamefont{{Shao}}},
  \bibinfo{author}{\bibfnamefont{A.}~\bibnamefont{{Taracchini}}},
  \bibinfo{author}{\bibfnamefont{A.}~\bibnamefont{{Buonanno}}},
  \bibinfo{author}{\bibfnamefont{S.}~\bibnamefont{{Babak}}},
  \bibinfo{author}{\bibfnamefont{I.~W.} \bibnamefont{{Harry}}},
  \bibinfo{author}{\bibfnamefont{I.}~\bibnamefont{{Hinder}}},
  \bibinfo{author}{\bibfnamefont{S.}~\bibnamefont{{Ossokine}}},
  \bibinfo{author}{\bibfnamefont{M.}~\bibnamefont{{P{\"u}rrer}}},
  \bibinfo{author}{\bibfnamefont{V.}~\bibnamefont{{Raymond}}},
  \bibnamefont{et~al.}, \bibinfo{journal}{\prd} \textbf{\bibinfo{volume}{95}},
  \bibinfo{eid}{044028} (\bibinfo{year}{2017}), \eprint{1611.03703}.

\bibitem[{\citenamefont{{Varma} et~al.}(2019)\citenamefont{{Varma}, {Field},
  {Scheel}, {Blackman}, {Gerosa}, {Stein}, {Kidder}, and
  {Pfeiffer}}}]{2019arXiv190509300V}
\bibinfo{author}{\bibfnamefont{V.}~\bibnamefont{{Varma}}},
  \bibinfo{author}{\bibfnamefont{S.~E.} \bibnamefont{{Field}}},
  \bibinfo{author}{\bibfnamefont{M.~A.} \bibnamefont{{Scheel}}},
  \bibinfo{author}{\bibfnamefont{J.}~\bibnamefont{{Blackman}}},
  \bibinfo{author}{\bibfnamefont{D.}~\bibnamefont{{Gerosa}}},
  \bibinfo{author}{\bibfnamefont{L.~C.} \bibnamefont{{Stein}}},
  \bibinfo{author}{\bibfnamefont{L.~E.} \bibnamefont{{Kidder}}},
  \bibnamefont{and} \bibinfo{author}{\bibfnamefont{H.~P.}
  \bibnamefont{{Pfeiffer}}}, \bibinfo{journal}{Available as arxiv:1905.9300}
  (\bibinfo{year}{2019}), \eprint{1905.09300}.

\bibitem[{\citenamefont{{Pratten} et~al.}(2020)\citenamefont{{Pratten},
  {Garc{\'\i}a-Quir{\'o}s}, {Colleoni}, {Ramos-Buades}, {Estell{\'e}s},
  {Mateu-Lucena}, {Jaume}, {Haney}, {Keitel}, {Thompson}
  et~al.}}]{gwastro-mergers-IMRPhenomXP}
\bibinfo{author}{\bibfnamefont{G.}~\bibnamefont{{Pratten}}},
  \bibinfo{author}{\bibfnamefont{C.}~\bibnamefont{{Garc{\'\i}a-Quir{\'o}s}}},
  \bibinfo{author}{\bibfnamefont{M.}~\bibnamefont{{Colleoni}}},
  \bibinfo{author}{\bibfnamefont{A.}~\bibnamefont{{Ramos-Buades}}},
  \bibinfo{author}{\bibfnamefont{H.}~\bibnamefont{{Estell{\'e}s}}},
  \bibinfo{author}{\bibfnamefont{M.}~\bibnamefont{{Mateu-Lucena}}},
  \bibinfo{author}{\bibfnamefont{R.}~\bibnamefont{{Jaume}}},
  \bibinfo{author}{\bibfnamefont{M.}~\bibnamefont{{Haney}}},
  \bibinfo{author}{\bibfnamefont{D.}~\bibnamefont{{Keitel}}},
  \bibinfo{author}{\bibfnamefont{J.~E.} \bibnamefont{{Thompson}}},
  \bibnamefont{et~al.}, \bibinfo{journal}{arXiv e-prints}
  \bibinfo{eid}{arXiv:2004.06503} (\bibinfo{year}{2020}), \eprint{2004.06503}.

\bibitem[{\citenamefont{{Ossokine} et~al.}(2020)\citenamefont{{Ossokine},
  {Buonanno}, {Marsat}, {Cotesta}, {Babak}, {Dietrich}, {Haas}, {Hinder},
  {Pfeiffer}, {P{\"u}rrer} et~al.}}]{2020PhRvD.102d4055O}
\bibinfo{author}{\bibfnamefont{S.}~\bibnamefont{{Ossokine}}},
  \bibinfo{author}{\bibfnamefont{A.}~\bibnamefont{{Buonanno}}},
  \bibinfo{author}{\bibfnamefont{S.}~\bibnamefont{{Marsat}}},
  \bibinfo{author}{\bibfnamefont{R.}~\bibnamefont{{Cotesta}}},
  \bibinfo{author}{\bibfnamefont{S.}~\bibnamefont{{Babak}}},
  \bibinfo{author}{\bibfnamefont{T.}~\bibnamefont{{Dietrich}}},
  \bibinfo{author}{\bibfnamefont{R.}~\bibnamefont{{Haas}}},
  \bibinfo{author}{\bibfnamefont{I.}~\bibnamefont{{Hinder}}},
  \bibinfo{author}{\bibfnamefont{H.~P.} \bibnamefont{{Pfeiffer}}},
  \bibinfo{author}{\bibfnamefont{M.}~\bibnamefont{{P{\"u}rrer}}},
  \bibnamefont{et~al.}, \bibinfo{journal}{\prd} \textbf{\bibinfo{volume}{102}},
  \bibinfo{eid}{044055} (\bibinfo{year}{2020}), \eprint{2004.09442}.

\bibitem[{\citenamefont{{Wysocki}
  et~al.}(2019{\natexlab{a}})\citenamefont{{Wysocki}, {Lange}, and
  {O'Shaughnessy}}}]{gwastro-PopulationReconstruct-Parametric-Wysocki2018}
\bibinfo{author}{\bibfnamefont{D.}~\bibnamefont{{Wysocki}}},
  \bibinfo{author}{\bibfnamefont{J.}~\bibnamefont{{Lange}}}, \bibnamefont{and}
  \bibinfo{author}{\bibfnamefont{R.}~\bibnamefont{{O'Shaughnessy}}},
  \bibinfo{journal}{\prd} \textbf{\bibinfo{volume}{100}}, \bibinfo{pages}{3012}
  (\bibinfo{year}{2019}{\natexlab{a}}),
  \urlprefix\url{https://arxiv.org/abs/1805.06442}.

\bibitem[{\citenamefont{Ashton and Khan}(2020)}]{Ashton:2019leq}
\bibinfo{author}{\bibfnamefont{G.}~\bibnamefont{Ashton}} \bibnamefont{and}
  \bibinfo{author}{\bibfnamefont{S.}~\bibnamefont{Khan}},
  \bibinfo{journal}{Phys. Rev. D} \textbf{\bibinfo{volume}{101}},
  \bibinfo{pages}{064037} (\bibinfo{year}{2020}), \eprint{1910.09138}.

\bibitem[{\citenamefont{Veitch et~al.}(2015)}]{Veitch:2014wba}
\bibinfo{author}{\bibfnamefont{J.}~\bibnamefont{Veitch}} \bibnamefont{et~al.},
  \bibinfo{journal}{Phys. Rev. D} \textbf{\bibinfo{volume}{91}},
  \bibinfo{pages}{042003} (\bibinfo{year}{2015}), \eprint{1409.7215}.

\bibitem[{\citenamefont{{Ashton} et~al.}(2019)\citenamefont{{Ashton},
  {H{\"u}bner}, {Lasky}, {Talbot}, {Ackley}, {Biscoveanu}, {Chu}, {Divakarla},
  {Easter}, {Goncharov} et~al.}}]{2019ApJS..241...27A}
\bibinfo{author}{\bibfnamefont{G.}~\bibnamefont{{Ashton}}},
  \bibinfo{author}{\bibfnamefont{M.}~\bibnamefont{{H{\"u}bner}}},
  \bibinfo{author}{\bibfnamefont{P.~D.} \bibnamefont{{Lasky}}},
  \bibinfo{author}{\bibfnamefont{C.}~\bibnamefont{{Talbot}}},
  \bibinfo{author}{\bibfnamefont{K.}~\bibnamefont{{Ackley}}},
  \bibinfo{author}{\bibfnamefont{S.}~\bibnamefont{{Biscoveanu}}},
  \bibinfo{author}{\bibfnamefont{Q.}~\bibnamefont{{Chu}}},
  \bibinfo{author}{\bibfnamefont{A.}~\bibnamefont{{Divakarla}}},
  \bibinfo{author}{\bibfnamefont{P.~J.} \bibnamefont{{Easter}}},
  \bibinfo{author}{\bibfnamefont{B.}~\bibnamefont{{Goncharov}}},
  \bibnamefont{et~al.}, \bibinfo{journal}{\apjs}
  \textbf{\bibinfo{volume}{241}}, \bibinfo{eid}{27} (\bibinfo{year}{2019}),
  \eprint{1811.02042}.

\bibitem[{\citenamefont{{Lange} et~al.}(2018)\citenamefont{{Lange},
  {O'Shaughnessy}, and {Rizzo}}}]{gwastro-PENR-RIFT}
\bibinfo{author}{\bibfnamefont{J.}~\bibnamefont{{Lange}}},
  \bibinfo{author}{\bibfnamefont{R.}~\bibnamefont{{O'Shaughnessy}}},
  \bibnamefont{and} \bibinfo{author}{\bibfnamefont{M.}~\bibnamefont{{Rizzo}}},
  \bibinfo{journal}{Submitted to PRD; available at arxiv:1805.10457}
  (\bibinfo{year}{2018}).

\bibitem[{\citenamefont{{Wysocki}
  et~al.}(2019{\natexlab{b}})\citenamefont{{Wysocki}, {O'Shaughnessy}, {Lange},
  and {Fang}}}]{gwastro-PENR-RIFT-GPU}
\bibinfo{author}{\bibfnamefont{D.}~\bibnamefont{{Wysocki}}},
  \bibinfo{author}{\bibfnamefont{R.}~\bibnamefont{{O'Shaughnessy}}},
  \bibinfo{author}{\bibfnamefont{J.}~\bibnamefont{{Lange}}}, \bibnamefont{and}
  \bibinfo{author}{\bibfnamefont{Y.-L.~L.} \bibnamefont{{Fang}}},
  \bibinfo{journal}{\prd} \textbf{\bibinfo{volume}{99}}, \bibinfo{eid}{084026}
  (\bibinfo{year}{2019}{\natexlab{b}}), \eprint{1902.04934}.

\bibitem[{\citenamefont{{Lange}}(2020)}]{gwastro-mergers-nr-LangePhD}
\bibinfo{author}{\bibfnamefont{J.}~\bibnamefont{{Lange}}},
  \emph{\bibinfo{title}{{RIFT'ing the Wave: Developing and applying an
  algorithm to infer properties gravitational wave sources}}}
  (\bibinfo{year}{2020}), \urlprefix\url{https://dcc.ligo.org/LIGO-P2000268}.

\bibitem[{\citenamefont{{Husa} et~al.}(2016)\citenamefont{{Husa}, {Khan},
  {Hannam}, {P{\"u}rrer}, {Ohme}, {Forteza}, and {Boh{\'e}}}}]{IMRDI}
\bibinfo{author}{\bibfnamefont{S.}~\bibnamefont{{Husa}}},
  \bibinfo{author}{\bibfnamefont{S.}~\bibnamefont{{Khan}}},
  \bibinfo{author}{\bibfnamefont{M.}~\bibnamefont{{Hannam}}},
  \bibinfo{author}{\bibfnamefont{M.}~\bibnamefont{{P{\"u}rrer}}},
  \bibinfo{author}{\bibfnamefont{F.}~\bibnamefont{{Ohme}}},
  \bibinfo{author}{\bibfnamefont{X.~J.} \bibnamefont{{Forteza}}},
  \bibnamefont{and}
  \bibinfo{author}{\bibfnamefont{A.}~\bibnamefont{{Boh{\'e}}}},
  \bibinfo{journal}{\prd} \textbf{\bibinfo{volume}{93}}, \bibinfo{eid}{044006}
  (\bibinfo{year}{2016}), \eprint{1508.07250}.

\bibitem[{\citenamefont{{Khan} et~al.}(2016{\natexlab{a}})\citenamefont{{Khan},
  {Husa}, {Hannam}, {Ohme}, {P{\"u}rrer}, {Forteza}, and {Boh{\'e}}}}]{IMRDII}
\bibinfo{author}{\bibfnamefont{S.}~\bibnamefont{{Khan}}},
  \bibinfo{author}{\bibfnamefont{S.}~\bibnamefont{{Husa}}},
  \bibinfo{author}{\bibfnamefont{M.}~\bibnamefont{{Hannam}}},
  \bibinfo{author}{\bibfnamefont{F.}~\bibnamefont{{Ohme}}},
  \bibinfo{author}{\bibfnamefont{M.}~\bibnamefont{{P{\"u}rrer}}},
  \bibinfo{author}{\bibfnamefont{X.~J.} \bibnamefont{{Forteza}}},
  \bibnamefont{and}
  \bibinfo{author}{\bibfnamefont{A.}~\bibnamefont{{Boh{\'e}}}},
  \bibinfo{journal}{\prd} \textbf{\bibinfo{volume}{93}}, \bibinfo{eid}{044007}
  (\bibinfo{year}{2016}{\natexlab{a}}), \eprint{1508.07253}.

\bibitem[{\citenamefont{{Taracchini} et~al.}(2012)\citenamefont{{Taracchini},
  {Pan}, {Buonanno}, {Barausse}, {Boyle}, {Chu}, {Lovelace}, {Pfeiffer}, and
  {Scheel}}}]{gw-astro-EOBspin-Tarrachini2012}
\bibinfo{author}{\bibfnamefont{A.}~\bibnamefont{{Taracchini}}},
  \bibinfo{author}{\bibfnamefont{Y.}~\bibnamefont{{Pan}}},
  \bibinfo{author}{\bibfnamefont{A.}~\bibnamefont{{Buonanno}}},
  \bibinfo{author}{\bibfnamefont{E.}~\bibnamefont{{Barausse}}},
  \bibinfo{author}{\bibfnamefont{M.}~\bibnamefont{{Boyle}}},
  \bibinfo{author}{\bibfnamefont{T.}~\bibnamefont{{Chu}}},
  \bibinfo{author}{\bibfnamefont{G.}~\bibnamefont{{Lovelace}}},
  \bibinfo{author}{\bibfnamefont{H.~P.} \bibnamefont{{Pfeiffer}}},
  \bibnamefont{and} \bibinfo{author}{\bibfnamefont{M.~A.}
  \bibnamefont{{Scheel}}}, \bibinfo{journal}{\prd}
  \textbf{\bibinfo{volume}{86}}, \bibinfo{eid}{024011} (\bibinfo{year}{2012}),
  \eprint{1202.0790}.

\bibitem[{\citenamefont{{Ajith} et~al.}(2007)\citenamefont{{Ajith}, {Babak},
  {Chen}, {Hewitson}, {Krishnan}, {Whelan}, {Br{\"u}gmann}, {Diener},
  {Gonzalez}, {Hannam} et~al.}}]{nr-Jena-nonspinning-templates2007}
\bibinfo{author}{\bibfnamefont{P.}~\bibnamefont{{Ajith}}},
  \bibinfo{author}{\bibfnamefont{S.}~\bibnamefont{{Babak}}},
  \bibinfo{author}{\bibfnamefont{Y.}~\bibnamefont{{Chen}}},
  \bibinfo{author}{\bibfnamefont{M.}~\bibnamefont{{Hewitson}}},
  \bibinfo{author}{\bibfnamefont{B.}~\bibnamefont{{Krishnan}}},
  \bibinfo{author}{\bibfnamefont{J.~T.} \bibnamefont{{Whelan}}},
  \bibinfo{author}{\bibfnamefont{B.}~\bibnamefont{{Br{\"u}gmann}}},
  \bibinfo{author}{\bibfnamefont{P.}~\bibnamefont{{Diener}}},
  \bibinfo{author}{\bibfnamefont{J.}~\bibnamefont{{Gonzalez}}},
  \bibinfo{author}{\bibfnamefont{M.}~\bibnamefont{{Hannam}}},
  \bibnamefont{et~al.}, \bibinfo{journal}{Classical and Quantum Gravity}
  \textbf{\bibinfo{volume}{24}}, \bibinfo{pages}{689} (\bibinfo{year}{2007}),
  \eprint{0704.3764}, \urlprefix\url{http://xxx.lanl.gov/abs/arxiv:0704.3764}.

\bibitem[{\citenamefont{{Santamar{\'{\i}}a}
  et~al.}(2010)\citenamefont{{Santamar{\'{\i}}a}, {Ohme}, {Ajith},
  {Br{\"u}gmann}, {Dorband}, {Hannam}, {Husa}, {M{\"o}sta}, {Pollney},
  {Reisswig} et~al.}}]{gwastro-nr-Phenom-Lucia2010}
\bibinfo{author}{\bibfnamefont{L.}~\bibnamefont{{Santamar{\'{\i}}a}}},
  \bibinfo{author}{\bibfnamefont{F.}~\bibnamefont{{Ohme}}},
  \bibinfo{author}{\bibfnamefont{P.}~\bibnamefont{{Ajith}}},
  \bibinfo{author}{\bibfnamefont{B.}~\bibnamefont{{Br{\"u}gmann}}},
  \bibinfo{author}{\bibfnamefont{N.}~\bibnamefont{{Dorband}}},
  \bibinfo{author}{\bibfnamefont{M.}~\bibnamefont{{Hannam}}},
  \bibinfo{author}{\bibfnamefont{S.}~\bibnamefont{{Husa}}},
  \bibinfo{author}{\bibfnamefont{P.}~\bibnamefont{{M{\"o}sta}}},
  \bibinfo{author}{\bibfnamefont{D.}~\bibnamefont{{Pollney}}},
  \bibinfo{author}{\bibfnamefont{C.}~\bibnamefont{{Reisswig}}},
  \bibnamefont{et~al.}, \bibinfo{journal}{\prd} \textbf{\bibinfo{volume}{82}},
  \bibinfo{pages}{064016} (\bibinfo{year}{2010}),
  \urlprefix\url{http://xxx.lanl.gov/abs/arXiv:1005.3306}.

\bibitem[{\citenamefont{{The LIGO Scientific Collaboration}
  et~al.}(2019)\citenamefont{{The LIGO Scientific Collaboration}, {The Virgo
  Collaboration}, {Abbott}, {Abbott}, {Abbott}, {Acernese}, {Ackley}, {Adams},
  {Adams}, {Addesso} et~al.}}]{LIGO-O2-Catalog}
\bibinfo{author}{\bibnamefont{{The LIGO Scientific Collaboration}}},
  \bibinfo{author}{\bibnamefont{{The Virgo Collaboration}}},
  \bibinfo{author}{\bibfnamefont{B.~P.} \bibnamefont{{Abbott}}},
  \bibinfo{author}{\bibfnamefont{R.}~\bibnamefont{{Abbott}}},
  \bibinfo{author}{\bibfnamefont{T.~D.} \bibnamefont{{Abbott}}},
  \bibinfo{author}{\bibfnamefont{F.}~\bibnamefont{{Acernese}}},
  \bibinfo{author}{\bibfnamefont{K.}~\bibnamefont{{Ackley}}},
  \bibinfo{author}{\bibfnamefont{C.}~\bibnamefont{{Adams}}},
  \bibinfo{author}{\bibfnamefont{T.}~\bibnamefont{{Adams}}},
  \bibinfo{author}{\bibfnamefont{P.}~\bibnamefont{{Addesso}}},
  \bibnamefont{et~al.}, \bibinfo{journal}{\prx} \textbf{\bibinfo{volume}{9}},
  \bibinfo{eid}{031040} (\bibinfo{year}{2019}).

\bibitem[{\citenamefont{{LIGO Scientific
  Collaboration}}(2018)}]{LIGO-aLIGODesign-Sensitivity-Updated}
\bibinfo{author}{\bibnamefont{{LIGO Scientific Collaboration}}}
  (\bibinfo{year}{2018}), \urlprefix\url{https://dcc.ligo.org/LIGO-T1800044}.

\bibitem[{\citenamefont{{Cook} et~al.}(2012)\citenamefont{{Cook}, {Gelman}, and
  {Rubin}}}]{mm-stats-PP}
\bibinfo{author}{\bibfnamefont{S.}~\bibnamefont{{Cook}}},
  \bibinfo{author}{\bibfnamefont{A.}~\bibnamefont{{Gelman}}}, \bibnamefont{and}
  \bibinfo{author}{\bibfnamefont{D.}~\bibnamefont{{Rubin}}},
  \bibinfo{journal}{Journal of Computational and Graphical Statistics}
  \textbf{\bibinfo{volume}{15}}, \bibinfo{pages}{675} (\bibinfo{year}{2012}),
  \urlprefix\url{https://www.tandfonline.com/doi/abs/10.1198/106186006X136976}.

\bibitem[{\citenamefont{{Lindblom} et~al.}(2008)\citenamefont{{Lindblom},
  {Owen}, and {Brown}}}]{gr-nr-WaveformErrorStandards-LBO-2008}
\bibinfo{author}{\bibfnamefont{L.}~\bibnamefont{{Lindblom}}},
  \bibinfo{author}{\bibfnamefont{B.~J.} \bibnamefont{{Owen}}},
  \bibnamefont{and} \bibinfo{author}{\bibfnamefont{D.~A.}
  \bibnamefont{{Brown}}}, \bibinfo{journal}{\prd}
  \textbf{\bibinfo{volume}{78}}, \bibinfo{pages}{124020}
  (\bibinfo{year}{2008}), \eprint{0809.3844},
  \urlprefix\url{http://xxx.lanl.gov/abs/arXiv:0809.3844}.

\bibitem[{\citenamefont{{Read} et~al.}(2009)\citenamefont{{Read}, {Markakis},
  {Shibata}, {Ury{\={u}}}, {Creighton}, and {Friedman}}}]{2009PhRvD..79l4033R}
\bibinfo{author}{\bibfnamefont{J.~S.} \bibnamefont{{Read}}},
  \bibinfo{author}{\bibfnamefont{C.}~\bibnamefont{{Markakis}}},
  \bibinfo{author}{\bibfnamefont{M.}~\bibnamefont{{Shibata}}},
  \bibinfo{author}{\bibfnamefont{K.}~\bibnamefont{{Ury{\={u}}}}},
  \bibinfo{author}{\bibfnamefont{J.~D.~E.} \bibnamefont{{Creighton}}},
  \bibnamefont{and} \bibinfo{author}{\bibfnamefont{J.~L.}
  \bibnamefont{{Friedman}}}, \bibinfo{journal}{\prd}
  \textbf{\bibinfo{volume}{79}}, \bibinfo{eid}{124033} (\bibinfo{year}{2009}),
  \eprint{0901.3258}.

\bibitem[{\citenamefont{{Lindblom} et~al.}(2010)\citenamefont{{Lindblom},
  {Baker}, and {Owen}}}]{2010PhRvD..82h4020L}
\bibinfo{author}{\bibfnamefont{L.}~\bibnamefont{{Lindblom}}},
  \bibinfo{author}{\bibfnamefont{J.~G.} \bibnamefont{{Baker}}},
  \bibnamefont{and} \bibinfo{author}{\bibfnamefont{B.~J.}
  \bibnamefont{{Owen}}}, \bibinfo{journal}{\prd} \textbf{\bibinfo{volume}{82}},
  \bibinfo{pages}{084020} (\bibinfo{year}{2010}), \eprint{1008.1803}.

\bibitem[{\citenamefont{{Cho} et~al.}(2013)\citenamefont{{Cho}, {Ochsner},
  {O'Shaughnessy}, {Kim}, and
  {Lee}}}]{gwastro-mergers-HeeSuk-FisherMatrixWithAmplitudeCorrections}
\bibinfo{author}{\bibfnamefont{H.}~\bibnamefont{{Cho}}},
  \bibinfo{author}{\bibfnamefont{E.}~\bibnamefont{{Ochsner}}},
  \bibinfo{author}{\bibfnamefont{R.}~\bibnamefont{{O'Shaughnessy}}},
  \bibinfo{author}{\bibfnamefont{C.}~\bibnamefont{{Kim}}}, \bibnamefont{and}
  \bibinfo{author}{\bibfnamefont{C.}~\bibnamefont{{Lee}}},
  \bibinfo{journal}{\prd} \textbf{\bibinfo{volume}{87}}, \bibinfo{pages}{02400}
  (\bibinfo{year}{2013}), \eprint{1209.4494},
  \urlprefix\url{http://xxx.lanl.gov/abs/arXiv:1209.4494}.

\bibitem[{\citenamefont{{Hannam} et~al.}(2010)\citenamefont{{Hannam}, {Husa},
  {Ohme}, and {Ajith}}}]{2010PhRvD..82l4052H}
\bibinfo{author}{\bibfnamefont{M.}~\bibnamefont{{Hannam}}},
  \bibinfo{author}{\bibfnamefont{S.}~\bibnamefont{{Husa}}},
  \bibinfo{author}{\bibfnamefont{F.}~\bibnamefont{{Ohme}}}, \bibnamefont{and}
  \bibinfo{author}{\bibfnamefont{P.}~\bibnamefont{{Ajith}}},
  \bibinfo{journal}{\prd} \textbf{\bibinfo{volume}{82}}, \bibinfo{eid}{124052}
  (\bibinfo{year}{2010}), \eprint{1008.2961}.

\bibitem[{\citenamefont{{Kumar} et~al.}(2016)\citenamefont{{Kumar}, {Chu},
  {Fong}, {Pfeiffer}, {Boyle}, {Hemberger}, {Kidder}, {Scheel}, and
  {Szilagyi}}}]{2016PhRvD..93j4050K}
\bibinfo{author}{\bibfnamefont{P.}~\bibnamefont{{Kumar}}},
  \bibinfo{author}{\bibfnamefont{T.}~\bibnamefont{{Chu}}},
  \bibinfo{author}{\bibfnamefont{H.}~\bibnamefont{{Fong}}},
  \bibinfo{author}{\bibfnamefont{H.~P.} \bibnamefont{{Pfeiffer}}},
  \bibinfo{author}{\bibfnamefont{M.}~\bibnamefont{{Boyle}}},
  \bibinfo{author}{\bibfnamefont{D.~A.} \bibnamefont{{Hemberger}}},
  \bibinfo{author}{\bibfnamefont{L.~E.} \bibnamefont{{Kidder}}},
  \bibinfo{author}{\bibfnamefont{M.~A.} \bibnamefont{{Scheel}}},
  \bibnamefont{and}
  \bibinfo{author}{\bibfnamefont{B.}~\bibnamefont{{Szilagyi}}},
  \bibinfo{journal}{\prd} \textbf{\bibinfo{volume}{93}}, \bibinfo{eid}{104050}
  (\bibinfo{year}{2016}), \eprint{1601.05396}.

\bibitem[{\citenamefont{{P{\"u}rrer} and {Haster}}(2020)}]{2020PhRvR...2b3151P}
\bibinfo{author}{\bibfnamefont{M.}~\bibnamefont{{P{\"u}rrer}}}
  \bibnamefont{and} \bibinfo{author}{\bibfnamefont{C.-J.}
  \bibnamefont{{Haster}}}, \bibinfo{journal}{Physical Review Research}
  \textbf{\bibinfo{volume}{2}}, \bibinfo{eid}{023151} (\bibinfo{year}{2020}),
  \eprint{1912.10055}.

\bibitem[{\citenamefont{{Khan} et~al.}(2016{\natexlab{b}})\citenamefont{{Khan},
  {Husa}, {Hannam}, {Ohme}, {P{\"u}rrer}, {Forteza}, and
  {Boh{\'e}}}}]{2016PhRvD..93d4007K}
\bibinfo{author}{\bibfnamefont{S.}~\bibnamefont{{Khan}}},
  \bibinfo{author}{\bibfnamefont{S.}~\bibnamefont{{Husa}}},
  \bibinfo{author}{\bibfnamefont{M.}~\bibnamefont{{Hannam}}},
  \bibinfo{author}{\bibfnamefont{F.}~\bibnamefont{{Ohme}}},
  \bibinfo{author}{\bibfnamefont{M.}~\bibnamefont{{P{\"u}rrer}}},
  \bibinfo{author}{\bibfnamefont{X.~J.} \bibnamefont{{Forteza}}},
  \bibnamefont{and}
  \bibinfo{author}{\bibfnamefont{A.}~\bibnamefont{{Boh{\'e}}}},
  \bibinfo{journal}{\prd} \textbf{\bibinfo{volume}{93}}, \bibinfo{eid}{044007}
  (\bibinfo{year}{2016}{\natexlab{b}}), \eprint{1508.07253}.

\bibitem[{\citenamefont{{Kumar} et~al.}(2014)\citenamefont{{Kumar},
  {MacDonald}, {Brown}, {Pfeiffer}, {Cannon}, {Boyle}, {Kidder}, {Mrou{\'e}},
  {Scheel}, {Szil{\'a}gyi} et~al.}}]{2014PhRvD..89d2002K}
\bibinfo{author}{\bibfnamefont{P.}~\bibnamefont{{Kumar}}},
  \bibinfo{author}{\bibfnamefont{I.}~\bibnamefont{{MacDonald}}},
  \bibinfo{author}{\bibfnamefont{D.~A.} \bibnamefont{{Brown}}},
  \bibinfo{author}{\bibfnamefont{H.~P.} \bibnamefont{{Pfeiffer}}},
  \bibinfo{author}{\bibfnamefont{K.}~\bibnamefont{{Cannon}}},
  \bibinfo{author}{\bibfnamefont{M.}~\bibnamefont{{Boyle}}},
  \bibinfo{author}{\bibfnamefont{L.~E.} \bibnamefont{{Kidder}}},
  \bibinfo{author}{\bibfnamefont{A.~H.} \bibnamefont{{Mrou{\'e}}}},
  \bibinfo{author}{\bibfnamefont{M.~A.} \bibnamefont{{Scheel}}},
  \bibinfo{author}{\bibfnamefont{B.}~\bibnamefont{{Szil{\'a}gyi}}},
  \bibnamefont{et~al.}, \bibinfo{journal}{\prd} \textbf{\bibinfo{volume}{89}},
  \bibinfo{eid}{042002} (\bibinfo{year}{2014}), \eprint{1310.7949}.

\bibitem[{\citenamefont{{Kumar} et~al.}(2015)\citenamefont{{Kumar}, {Barkett},
  {Bhagwat}, {Afshari}, {Brown}, {Lovelace}, {Scheel}, and
  {Szil{\'a}gyi}}}]{2015PhRvD..92j2001K}
\bibinfo{author}{\bibfnamefont{P.}~\bibnamefont{{Kumar}}},
  \bibinfo{author}{\bibfnamefont{K.}~\bibnamefont{{Barkett}}},
  \bibinfo{author}{\bibfnamefont{S.}~\bibnamefont{{Bhagwat}}},
  \bibinfo{author}{\bibfnamefont{N.}~\bibnamefont{{Afshari}}},
  \bibinfo{author}{\bibfnamefont{D.~A.} \bibnamefont{{Brown}}},
  \bibinfo{author}{\bibfnamefont{G.}~\bibnamefont{{Lovelace}}},
  \bibinfo{author}{\bibfnamefont{M.~A.} \bibnamefont{{Scheel}}},
  \bibnamefont{and}
  \bibinfo{author}{\bibfnamefont{B.}~\bibnamefont{{Szil{\'a}gyi}}},
  \bibinfo{journal}{\prd} \textbf{\bibinfo{volume}{92}}, \bibinfo{eid}{102001}
  (\bibinfo{year}{2015}), \eprint{1507.00103}.

\bibitem[{\citenamefont{{Chua} et~al.}(2020)\citenamefont{{Chua}, {Korsakova},
  {Moore}, {Gair}, and {Babak}}}]{2020PhRvD.101d4027C}
\bibinfo{author}{\bibfnamefont{A.~J.~K.} \bibnamefont{{Chua}}},
  \bibinfo{author}{\bibfnamefont{N.}~\bibnamefont{{Korsakova}}},
  \bibinfo{author}{\bibfnamefont{C.~J.} \bibnamefont{{Moore}}},
  \bibinfo{author}{\bibfnamefont{J.~R.} \bibnamefont{{Gair}}},
  \bibnamefont{and} \bibinfo{author}{\bibfnamefont{S.}~\bibnamefont{{Babak}}},
  \bibinfo{journal}{\prd} \textbf{\bibinfo{volume}{101}}, \bibinfo{eid}{044027}
  (\bibinfo{year}{2020}), \eprint{1912.11543}.

\end{thebibliography}
\end{document}